\documentclass[twocolumn]{aastex61}

\usepackage{booktabs}
\usepackage{multirow}
\usepackage{graphicx}
\usepackage{natbib}
\usepackage{gensymb}
\usepackage{txfonts}

\newcommand{\tex}{T$_{ex}$ }
\newcommand{\tk}{T$_K$ }
\newcommand{\vc}{$V_{lsr}$ }
\newcommand{\sig}{$\sigma_V$ }
\newcommand{\col}{N(NH$_3$) }
\newcommand{\xam}{X(NH$_3$) }
\newcommand{\am}{NH$_3$ }
\newcommand{\kms}{km$\,$s$^{-1}$ }

\newcommand{\tdust}{T$_{dust}$ }

\submitjournal{ApJ}

\shorttitle{The GAS Survey: Unveiling the Dynamics of Barnard 59}
\shortauthors{Redaelli et al.}
\begin{document}

\title{The Green Bank Ammonia Survey: Unveiling the Dynamics of the Barnard 59 star-forming Clump}

\correspondingauthor{E. Redaelli}
\email{eredaelli@mpe.mpg.de}

\author{E. Redaelli}
\affiliation{Max Planck Institut f\"ur Extraterrestrische Physik, Giessenbachstrasse 1, D-85748, Garching, Germany}

\author{F. O. Alves}
\affiliation{Max Planck Institut f\"ur Extraterrestrische Physik, Giessenbachstrasse 1, D-85748, Garching, Germany}

\author{P. Caselli}
\affiliation{Max Planck Institut f\"ur Extraterrestrische Physik, Giessenbachstrasse 1, D-85748, Garching, Germany}

\author{J. E. Pineda}
\affiliation{Max Planck Institut f\"ur Extraterrestrische Physik, Giessenbachstrasse 1, D-85748, Garching, Germany}

\author{R. K. Friesen}
\affiliation{Dunlap Institute for Astronomy and Astrophysics, University of Toronto, 50 St. George St., Toronto, ON, M5S 3H4, Canada}

\author{A. Chac\'on-Tanarro}
\affiliation{Max Planck Institut f\"ur Extraterrestrische Physik, Giessenbachstrasse 1, D-85748, Garching, Germany}

\author{C. D. Matzner}
\affiliation{Department of Astronomy \& Astrophysics, University of Toronto, 50 St. George Street, Toronto, Ontario, Canada M5S 3H4}

\author{A. Ginsburg}
\affiliation{National Radio Astronomy Observatory, Socorro, NM 87801, USA}

\author{E. Rosolowsky}
\affiliation{Department of Physics, University of Alberta, Edmonton, AB, Canada}

\author{J. Keown}
\affiliation{Department of Physics and Astronomy, University of Victoria, Victoria, BC, V8P 5C2, Canada}

\author{S. S. R. Offner}
\affiliation{Department of Astronomy, University of Massachusetts, Amherst, MA 01003, USA}

\author{J. Di Francesco}
\affiliation{Department of Physics and Astronomy, University of Victoria, Victoria, BC, V8P 5C2, Canada}
\affiliation{NRC Herzberg Astronomy and Astrophysics, 5071 West Saanich Road, Victoria, BC, V9E 2E7, Canada}

\author{H. Kirk}
\affiliation{NRC Herzberg Astronomy and Astrophysics, 5071 West Saanich Road, Victoria, BC, V9E 2E7, Canada}

\author{P. C. Myers}
\affiliation{Harvard-Smithsonian Center for Astrophysics, 60 Garden St., Cambridge, MA 02138, USA}

\author{A. Hacar}
\affiliation{Leiden Observatory, Leiden University, P.O. Box 9513, 2300-RA Leiden, The Netherlands}
\affiliation{Institute for Astrophysics, University of Vienna, Türkenschanzstrasse 17, 1180 Vienna, Austria}

\author{A. Cimatti}
\affiliation{Dipartimento di Fisica e Astronomia, Universit\`a di Bologna, Via Piero Gobetti 93/2, 40129 Bologna, Italy}

\author{H. H. Chen}
\affiliation{Harvard-Smithsonian Center for Astrophysics, 60 Garden St., Cambridge, MA 02138, USA}

\author{M. C. Chen}
\affiliation{Department of Physics and Astronomy, University of Victoria, Victoria, BC, V8P 5C2, Canada}

\author{Y. M. Seo}
\affiliation{Jet Propulsion Laboratory, NASA, 4800 Oak Grove Dr., Pasadena, CA 91109, USA}

\author{K. I. Lee}
\affiliation{Harvard-Smithsonian Center for Astrophysics, 60 Garden St., Cambridge, MA 02138, USA}

\begin{abstract}
 Understanding the early stages of star formation is a research field of ongoing development, both theoretically and observationally. In this context, molecular data have been continuously providing observational constraints on the gas dynamics at different excitation conditions and depths in the sources. We have investigated the Barnard 59 core, the only active site of star formation in the Pipe Nebula, to achieve a comprehensive view of the kinematic properties of the source. These information were derived by simultaneously fitting ammonia inversion transition lines (1,1) and (2,2). Our analysis unveils the imprint of protostellar feedback, such as increasing line widths, temperature and turbulent motions in our molecular data. Combined with  complementary observations of dust thermal emission, we estimate that the core is gravitationally bound following a virial analysis. If the core is not contracting, another source of internal pressure, most likely the magnetic field, is supporting it against gravitational collapse and limits its star formation efficiency.
\end{abstract}


\section{Introduction} 
Low-mass star formation occurs through the fragmentation of molecular clouds into smaller and colder dense cores, where protostars can possibly form \citep{Andre00}. The efficiency for the whole process is low (e.g., usually less than $10 \, \%$ of the cloud mass is converted into stars, see \citealt{Evans09}) due to multiple reasons: cores can be transient objects because, for instance, they are not massive enough to collapse, or the feedback from the first protostars can disperse the gas before it can form new stars. Planck polarimetric observations \citep{PlanckXXXV} have revealed that molecular clouds are often magnetized, and that the magnetic fields that thread their filamentary structure \citep{Andre10} are usually aligned perpendicular to the filaments' long axes, especially at higher column density (see for instance \citealt{Palmeirim13} for the Taurus molecular cloud). In this configuration, magnetic fields can regulate the contraction of the structures more efficiently than interstellar turbulence, depending on the level of turbulence with respect to the magnetic field (e.g. \citealt{Nakamura08, Seifried15}). To better constrain the theoretical models, detailed observations of recent star-forming regions are needed. In particular, molecular data of rotational transitions, from which the source density, temperature, velocity and other parameters can be inferred, can help enlighten many aspects of the star formation that still lack a complete explanation, such as the effects of  feedback from young sources on the parental cloud or the role of the magnetic field. \par
The Pipe Nebula, a filamentary molecular cloud located at $\approx 145 \,$pc \citep{AlvesFranco07}, represents an ideal environment to study the early stages of star formation. The cloud is known to exhibit a very low star formation efficiency ($\approx 0.06 \%$, \citealt{Forbrich09}). Despite its reservoir of  cold gas ($M_{tot} \approx 10^4 \, M_{\odot}$, \citealt{Onishi99}, \citealt{Lombardi06}), only its extreme northwest end (Barnard 59, hereafter B59) shows signs of recent star-forming activity, as indicated by the detection of several young stellar objects (YSOs) by the Spitzer \textit{Cores to Disks} (c2d) survey \citep{Brooke07}. Furthermore, some of these YSOs are driving outflows, which inject enough energy in the cloud to sustain the turbulence level of the gas at sub-pc scales and to slow down the gravitational collapse \citep{DuarteCabral12}. This hypothesis is consistent with the monolithic dust distribution profile reported by \cite{RomanZuniga12}, which implies a lack of  further fragmentation. It is not clear why B59 is the only active region in a cloud with numerous prestellar cores (total of 159, \citealt{Lada08}). The answer can possibly be found in the morphology of the magnetic field, which is less uniform around B59 in comparison with the rest of the cloud \citep{Franco10}. These authors suggest that the locally disorganized field  may have allowed the core to form by turbulent compression. Moreover, \cite{Peretto12} suggested that a large-scale external compression, possibly due to stellar winds from the close Sco OB2 association, may have shaped the filamentary structure of the cloud, and it is interesting to notice that this association is closer to the western part of the Pipe, where B59 is located. A deeper analysis of the kinematics of B59 will help us understand the star forming process at core scales, and also determine its feedback on the parental cloud. \par
Ammonia (NH$_3$) emission has been an important tool for studying the coldest phases of the interstellar medium \citep{BensonMyers89}. Its rotational states with $J=K$ are split into doublets due to the ability of the nitrogen atom to tunnel quantum-mechanically through the plane of H atoms, giving rise to  the so called inversion transitions \citep{HoTownes83}, labeled as ($J$,$J$). They are characterized by a critical density $n_{cr}$\footnote{ $n_{cr}$ is computed as the ratio between the Einstein coefficient of spontaneous emission and the sum over the collisional rates.} $ \approx 10^{3-4} \,$cm$^{-3}$ \citep{Shirley15}.  In addition, the fact that ammonia does not appear to freeze-out significantly onto dust grains even at very low temperature (e.g. \citealt{Tafalla02}), makes \am one of the best probes for dense ($n \gtrsim 10^{4} \,$cm$^{-3}$) and cold ($T \lesssim 10 \,$K) environments. Spectral fitting of several ammonia inversion transitions therefore can provide crucial information about the source properties, such as its kinematic structure, density and temperature. \par
We used ammonia (1,1) and (2,2) data to investigate the kinematics of Barnard 59 and to analyze the protostellar feedback on the core (in Sec. \ref{ProtoFeed}). In Sec. \ref{CompHersch}, we combined the NH$_3$-derived information with dust thermal emission data at far infrared wavelengths, to compare the gas and the dust properties. We also performed a simple virial analysis, illustrated in Sec. \ref{VirialAna}, to investigate the core dynamical state.

\section{Observations}
The \am (1,1) and (2,2) inversion transitions in B59 were observed between 2015 and 2016 with the Green Bank Telescope (GBT) as part of a Large Program, the Green Bank Ammonia Survey (GAS - \citealt{FriesenPineda17}), whose aim is to study dense gas in several molecular clouds in the Gould Belt using ammonia and other molecular tracers. The GAS project used the K-band Focal Plane Array (KFPA) receiver combined with the VErsatile GBT Astronomical Spectrometer (VEGAS) backend in configuration Mode 20, which provides eight spectral windows for each of the seven KFPA beams with 23.44 MHz bandwidth and 4096 channels, corresponding to a spectral resolution of 5.72 KHz ($\approx 0.07\,$\kms at $\nu = 23 \,$GHz). Two spectral windows are centered at the ammonia (1,1) and (2,2) rest frequencies, which are respectively $\nu_{11} = 23694.4955 \,$MHz and $\nu_{22} = 23722.6336\,$MHz. More lines were observed, and will be presented in future papers. The chosen setup allows for an in-band frequency switching with a frequency shift of $4.11\,$MHz. At $23.7 \,$GHz, the KFPA beam size is FWHM $\approx 31.8''$, which at the distance of $145\,$pc is equivalent to a spatial resolution of $0.02\,$pc. The Moon and Jupiter were used as flux calibrators. Since the Moon's angular size is big enough to cover the entire KFPA sky footprint, it was used to determine the relative beam gains. GBTIDL\footnote{\url{http://gbtidl.nrao.edu/}} was used to read the raw data into sdfits format and to compute the relative beam gains. The data reduction was performed using Python scripts (available on GitHub\footnote{\url{https://github.com/GBTAmmoniaSurvey/GAS}}), which calls the GBT KFPA pipeline \citep{Masters11} for the calibration in main beam temperature (T$_{MB}$). Then the imaging was performed with custom GAS routines. The maps were acquired in On-The-Fly (OTF) mode across $10' \times 10'$ tiles. The observations were gridded according to \cite{Mangum07}, using a pixel size of $1/3$ of the beam FWHM at the \am (1,1) frequency. For further details on the observing setup or on the data reduction, please refer to \cite{FriesenPineda17}. Barnard 59 GAS data belong to the second data release (DR2) of the survey, and thus were not published in \cite{FriesenPineda17}.
\subsection{Results}
\begin{figure} [!h]
   \centering
   \includegraphics[width=  \hsize]{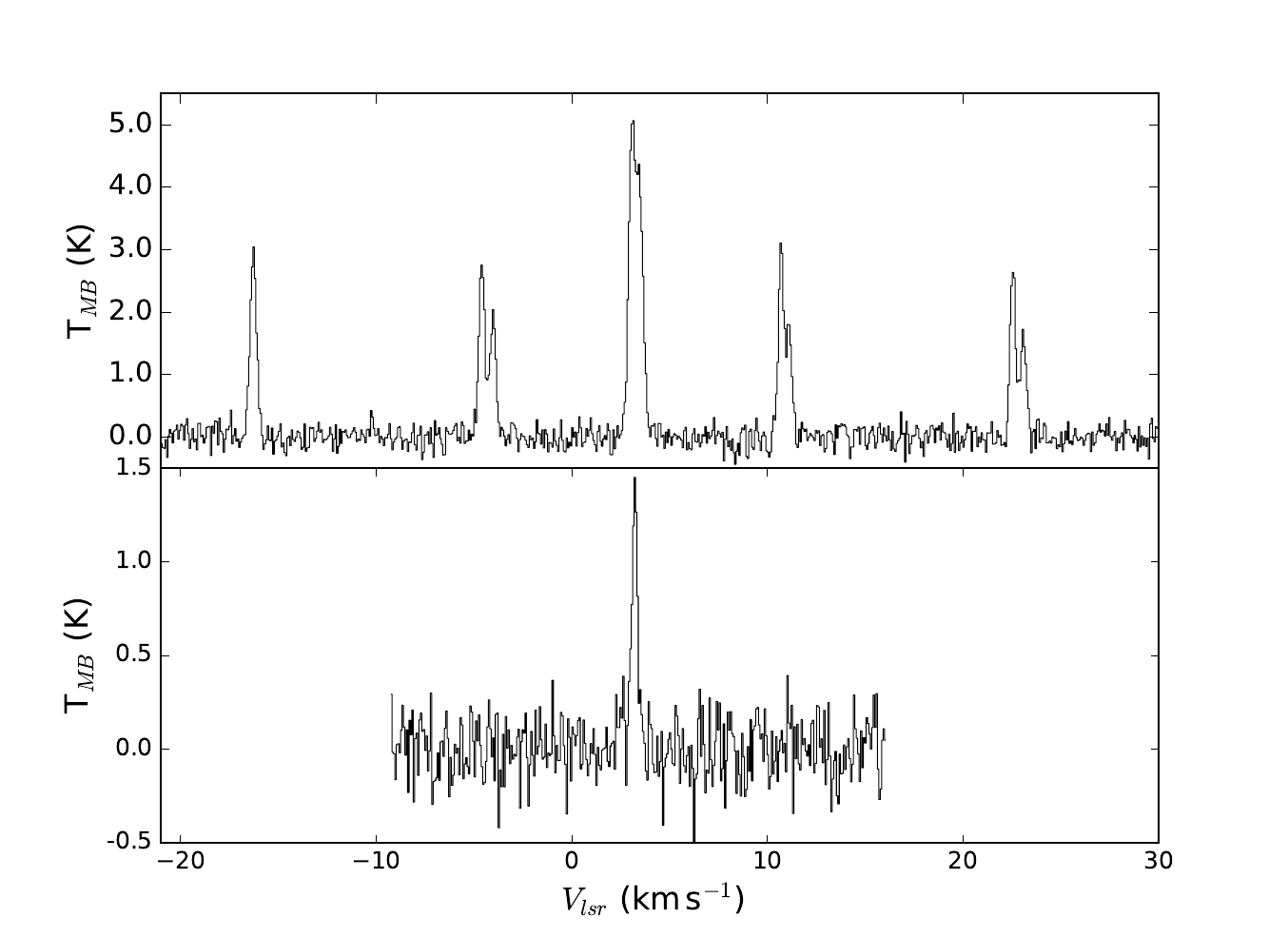}
   \caption{ \am (1,1) (upper panel) and (2,2) (lower panel) spectra from the brightest pixel (R.A.=$17^h11^m20^s.03$, Dec. = $-27^{\degree}26'27''.1$ (J2000)). The hyperfine structure is clearly visible for the (1,1), while for the (2,2) only the main group is detectable because the satellite components are too faint.}
  \label{spectra}
   \end{figure}

\begin{figure} [!h]
   \centering
   \includegraphics[width= \hsize]{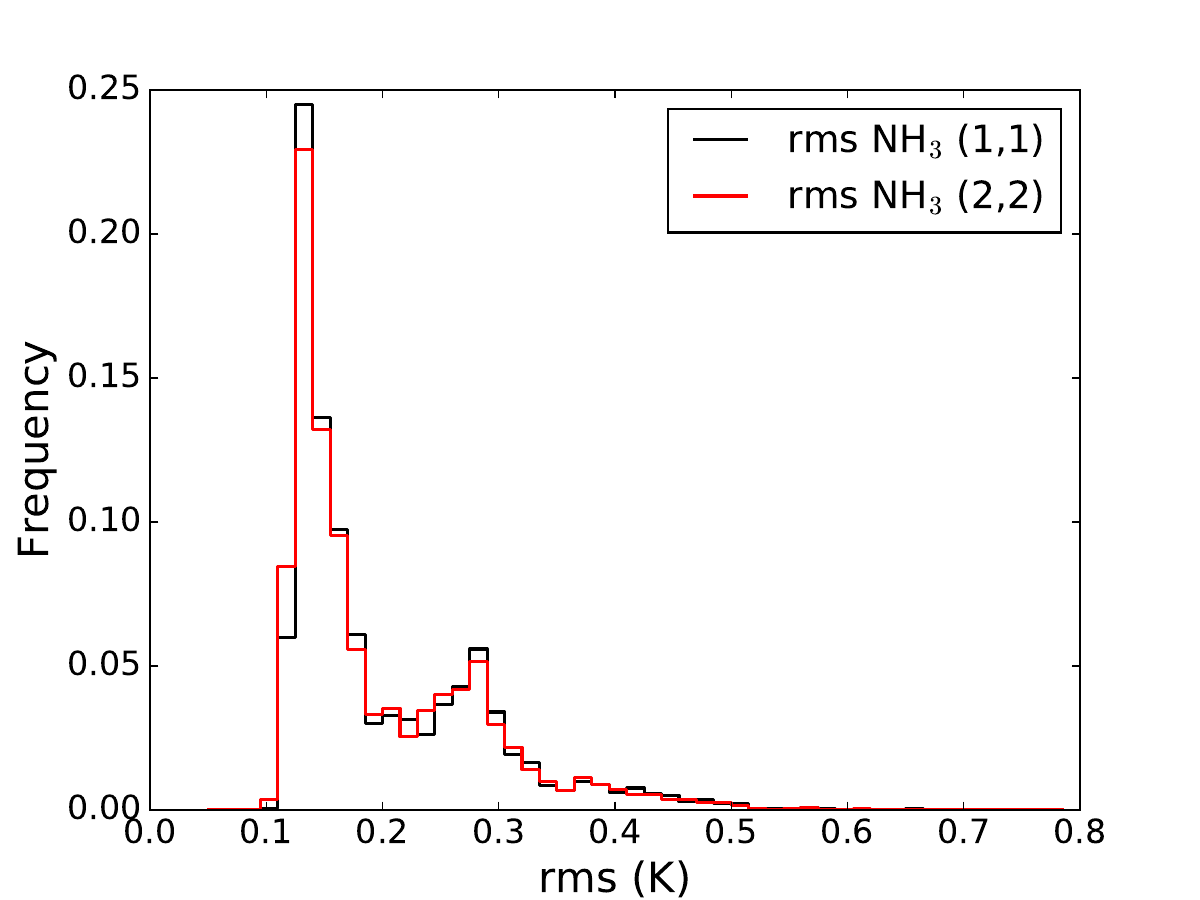}
   \caption{Frequency histogram of rms per channel for the (1,1) and the (2,2) lines in black and red, respectively. The data were binned with a spacing of $0.015\,$K.}
  \label{RMS}
   \end{figure}

The \am (1,1) and (2,2) spectra from the brightest pixel are shown in Figure \ref{spectra}.
 The rms noise map was determined by taking the intensity standard deviation for each pixel along the spectral axis for all line-free channels. Figure \ref{RMS} shows histograms of the rms distribution. For both lines a double peaked distribution is present, with the first maximum at rms$\; \approx 0.13\,$K. The second peak at higher values is due to pixels at the map's edges, where the on-source time is lower. If we exclude these noisy edges, the average rms value is $\approx 0.15\,$K per channel for both lines. This value is higher than the target level expected by the GAS survey ($\approx 0.10\,$K), mainly due to the fact that Barnard 59 is mostly at low elevation at the Green Bank latitude (maximum elevation: $\approx 25 \degree $). 
   
\subsection{Smoothing and Regridding \label{Smoo}}
The detection of at least two inversion transitions is necessary to determine the kinetic temperature and column density of  \am (see Sec. \ref{Flags} and Appendix \ref{NH3app}). The \am (2,2) line in most pixels is undetected, except for a small region in the central part of the dense core (less than $5\,$arcmin$^2$). To improve the signal-to-noise ratio (SNR) of the observations we smoothed the cubes to a resolution of FWHM$= 40''$. The regridded spectral cubes preserve the original number of three pixels per beam. Therefore, the new pixel size was computed following:
\begin{equation}
pxl\_sz_{new} =  \frac{\text{FWHM}_{new} }{3} \approx 13.33''.
\end{equation}
The regridding was performed using the python FITS\_tools\footnote{FITS\_tools is available at \url{https://github.com/keflavich/FITS_tools}} package.

\subsection{Ancillary data sets}
We used the dust emission data obtained with Herschel to derive the molecular hydrogen column density and the dust temperature in B59 (see section \ref{CompHersch}). We used the SPIRE maps of the Pipe Nebula at 250$\, \mu$m, 350$\, \mu$m and 500$\, \mu$m available in the Herschel Science Archive (Observational ID: 1342216013). We have selected the level 3 maps, i.e. the data are already calibrated and the maps are mosaiced. The angular resolutions of those data are approximately $20''$ ($250\, \mu$m), $27''$ ($350\, \mu$m) and $38''$ ($500\, \mu$m).\par
In addition, we used the LABOCA dust thermal emission map acquired with APEX telescope at $870\, \mu$m, to compute the total mass of the core. These data have a spatial resolution of $21.2''$. They are publicly available in the APEX database. 

\section{Data Analysis}

\subsection{Spectral fitting procedure}
The \am transitions have been analyzed to derive five line parameters: excitation temperature T$_{ex}$, kinetic temperature T$_{K}$, centroid velocity $V_{lsr}$, velocity dispersion $\sigma_V$, and ammonia total column density N(NH$_3$). The theoretical bases for this technique are explained in Appendix \ref{NH3app}. The spectra were fitted using these variables as free parameters. Our approach consists of a non-linear gradient descent regression, performed using the Python dedicated library PYSPECKIT \footnote{Available at \url{https://github.com/pyspeckit/pyspeckit.git}.} \citep{Ginsburg11}. The implemented code, available on the GAS GitHub repository, fits both lines simultaneously pixel by pixel starting from the position with the highest SNR and using their best fit values as guesses for nearby pixels. Table \ref{IniGues} summarizes the choices made for both initial guesses and parameter ranges, set to prevent unphysical results. The lower limit in velocity dispersion is set to be larger than the spectral resolution (FWHM$\; = 0.07\,$\kms corresponds to $\sigma \approx 0.03\,$km$\,$s$^{-1}$). The centroid velocity range is derived from the values obtained with C$^{18}$O in \cite{DuarteCabral12}. We masked pixels with low SNR based on the rms map and the peak value in the main beam brightness temperature of the \am (1,1) line (T$_{peak}$), according to SNR$_{peak} = $T$_{peak}/\text{rms} > 3.0$. The ortho-to-para ratio (OPR) has been fixed on OPR$\; = 1.0$, since we do not have any detection of ortho-\am transitions\footnote{Note that this is different in the approach (but not in the results) from the method used in \citep{FriesenPineda17}. There the OPR is set to 0 (no ortho states) and then the derived column density is doubled to obtain the total \col.}. The uncertainties for each parameter are estimated from the diagonal of the covariance matrix returned by the least-square fitter.

\begin{deluxetable}{ccc}
\tablewidth{0.0pt}
\tablecaption{Initial guesses for the fit. \label{IniGues}}
\tablecolumns{3}
\tablehead{ \colhead{Parameter} & \colhead{Initial guess} & \colhead{Value range}}
\startdata
$\log$[N(NH$_3$)/cm$^{-2}]$ &  $14.5$  & $ [12.0;17.0]$ \\
 \tk (K)  &$12.0$ & $[5.0; -]$ \\
 \tex (K) & $8.0$ & $[2.8; -]$ \\
 \vc (km$\,$s$^{-1}$)&  $3.37$\tablenotemark{a} & $[2.0;4.0]$ \\
  \sig (km$\,$s$^{-1}$)&  $0.11$\tablenotemark{a}  &$[0.04; - ]$ \\
  OPR &$1.0$ & fixed \\
 \enddata
 \tablenotetext{a}{\vc and \sig values derived from the calculation of moment 1 and moment 2 on the (1,1) cube.}
\end{deluxetable}

\subsection{Data masking \label{Flags}}
\begin{figure} [!h]
   \centering
   \includegraphics[width=\hsize]{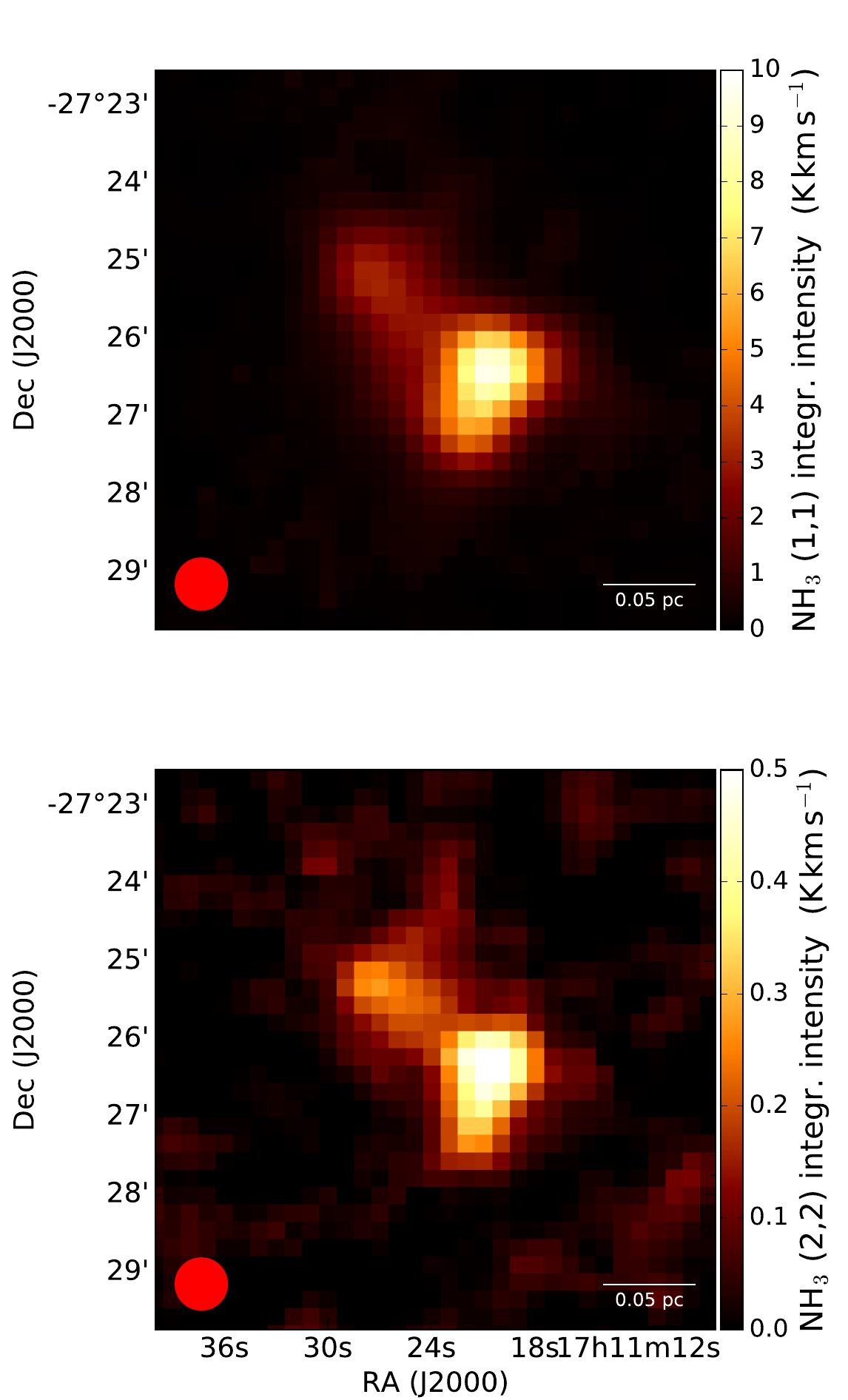}
   \caption{Integrated intensity maps of the \am (1,1)  (upper panel) and (2,2) (lower panel) lines, obtained with the method described in the  text (Sec. \ref{Flags}). The mean rms on source are $0.06\,$K$\,$\kms and $0.03 \,$K$\,$\kms respectively. Here and in all the following figures beam and scale bar are respectively shown in the bottom left and right corners.}
  \label{mom011}
   \end{figure}

The initial threshold imposed on SNR$_{peak}$ is sensitive to the presence of noise spikes in the spectra. Therefore, some pixels still present poorly constrained fitting results. To mask these, we took into account two further parameters: their uncertainties evaluated by the fitting procedure and the SNR in integrated intensity. The latter is calculated using the best fit spectra previously obtained to select the correct spectral windows for the integration. This approach is better than using a fixed spectral range for the whole map, since it takes into account velocity gradients. We flag channels with signal lower than $0.0125\,$K, and sum over the remaining ones. For those pixels where no line fit model is available, or at the positions where the fit is poorly constrained (which is evaluated taking into account the uncertainty on velocity dispersion), we adopt a fixed spectral window based on the mean \vc and \sig across the source.  At each pixel, the integrated intensity uncertainty is calculated through $\epsilon =  \text{rms} \, \Delta V \, \sqrt{N}$, where $\Delta V$ is the channel width and $N$ the number of channels involved in the integration. Figure \ref{mom011} shows the resulting integrated intensity maps for the (1,1) and (2,2) transitions. \par
Table \ref{dataflag} summarizes the criteria adopted to flag the data. Note that for T$_{ex}$, $V_{lsr}$ and $\sigma_V$, which are derived from the detection of (1,1) transition only, we chose a threshold on SNR in (1,1) integrated intensity. On the other hand, \tk and \col depend on the ratio between the (1,1) and (2,2) lines and therefore the additional SNR$_{22} > 3.0  $ condition is required. This approach differs from the one used in \citep{FriesenPineda17} in the sense that we also take into account each parameter uncertainties in the masking, and not only the integrated intensity SNR.

\begin{deluxetable*} {ccccc}
\tablecaption{Pixels masking criteria for the different free parameters. \label{dataflag}}
\tablehead{\colhead{\tk} & \colhead{\tex} & \colhead{\col} & \colhead{\sig} & \colhead{\vc}}
\startdata 
\hline
SNR$_{22} > 3.0  $ & SNR$_{11} > 3.0  $ & SNR$_{22} > 3.0  $ & SNR$_{11} > 3.0  $ & SNR$_{11} > 3.0  $ \\
\tk $ > 5.0 \, $K  &  \tex $ > 2.8 \, $K    &\tk $ > 5.0 \, $K  & $ \epsilon_{\sigma_V} <0.1 \,  $\kms & $ \epsilon_{\text{V}_{lsr}} <0.1 \,  $\kms \\
$ \epsilon_{\text{T}_K} < 3.0 \,$K &  $\epsilon_{\text{T}_{ex}} < 3.0 \, $K & $ \epsilon_{\text{T}_K} < 3.0 \,$K & & \\
       &		&  \tex $ > 2.8 \, $K  & & \\ 	
\enddata
\end{deluxetable*}


\subsection{Fitting results \label{ParaMaps}}
\begin{figure*} [h]
   \centering
   \includegraphics[width=   \hsize]{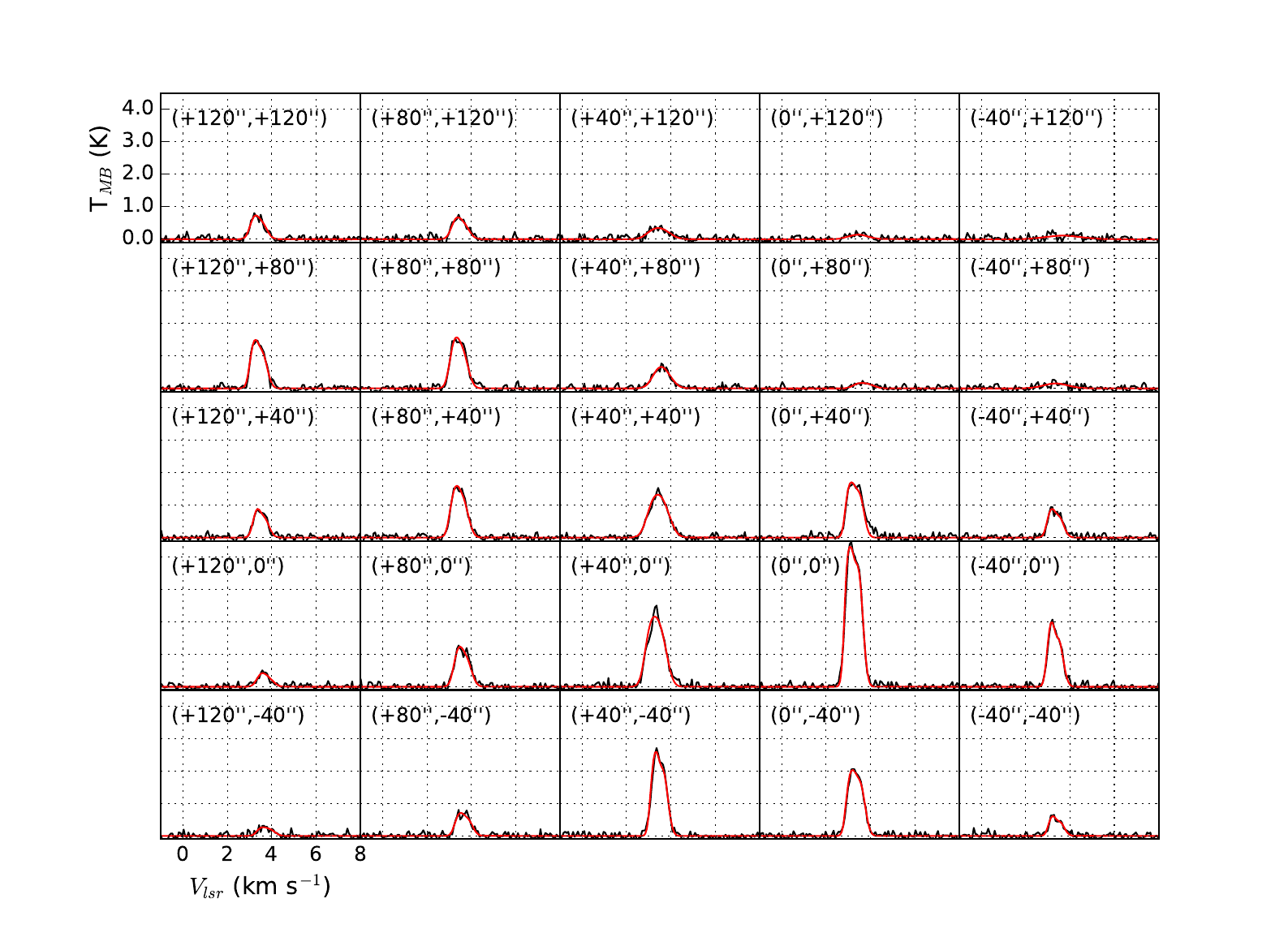}
   \caption{\am (1,1) observed spectra (black) and best-fit model (red) at 25 different positions. Only the main component is shown. In each panel the offset from the position R.A.=$17^h11^m19^s.99$, Dec. = $-27^{\degree}26'27''.3$ (J2000) is indicated as $(\Delta$R.A., $\Delta$Dec.) in arcsec.}
  \label{Spectra}
   \end{figure*}

Figure \ref{Spectra} shows the main component of the \am (1,1) spectra at different positions across B59, overlaid with the obtained best fit model. The positions are separated by a smoothed beam size. The overall quality of the fit is good. Across the entire source, we do not see evident signs of the presence of multiple velocity components along the line of sight.

\begin{figure} [!h]
   \centering
   \includegraphics[width= \hsize]{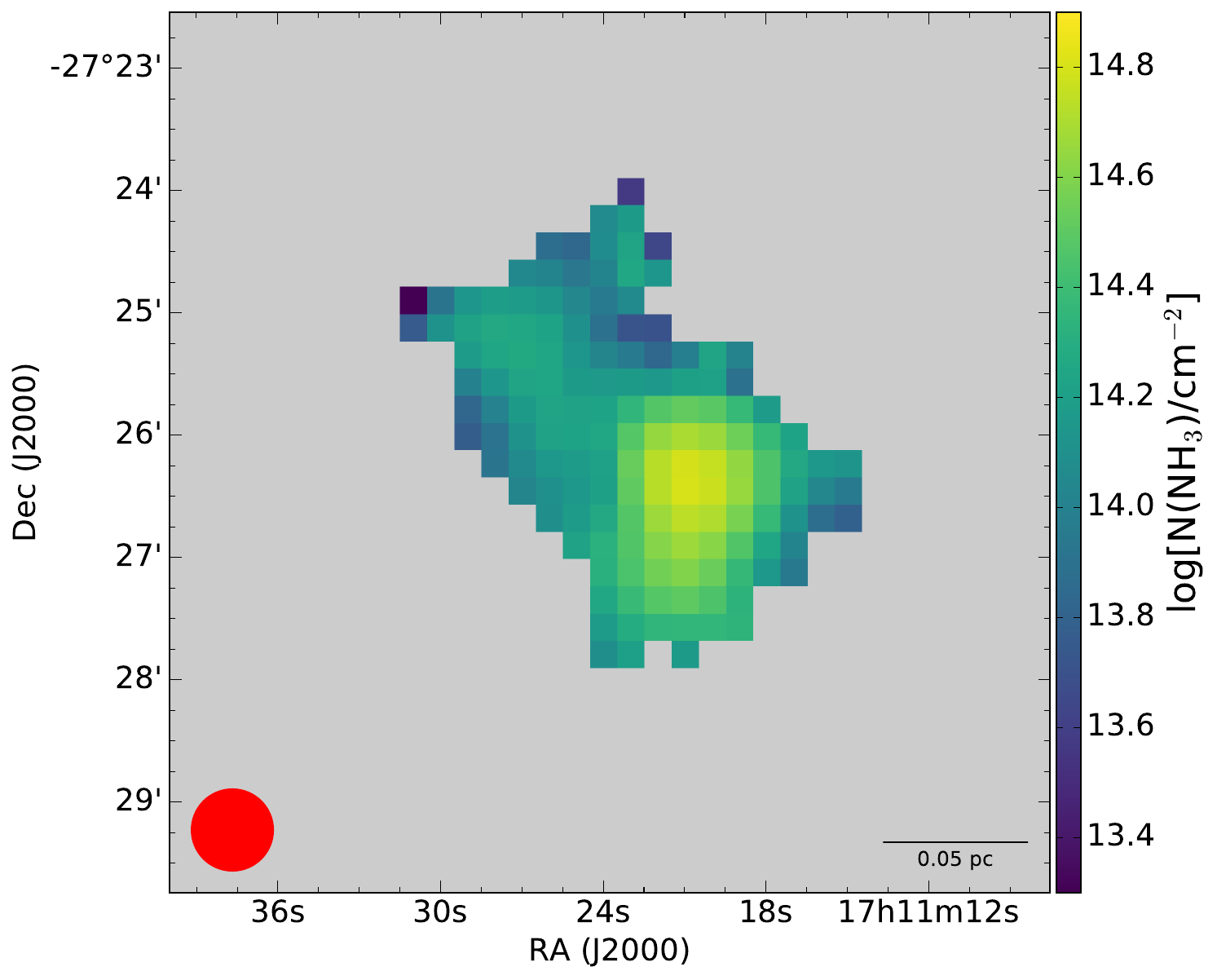}
   \caption{Ammonia column density map obtained in Barnard 59. The median uncertainty obtained is $0.04$.}
  \label{colden}
   \end{figure}

 The ammonia column density map, shown in Figure \ref{colden}, has a denser region in the south-west part of the source, with a peak of $\log[$N(NH$_3$)]$_{peak} = (14.80 \pm 0.01) \,$\footnote{Throughout the whole paper, column densities are expressed in unity of $\log_{10}[$cm$^{-2}]$, unless otherwise stated.} at R.A.=$17^h11^m21^s.00$, Dec. = $-27^{\degree}26'27''.3$ (J2000). A local increase of the \col is also present toward the NE direction, with a peak of $\log[$N(NH$_3$)]$_{peak} = (14.27 \pm 0.03)$ at R.A.=$17^h11^m27^s.00$, Dec. = $-27^{\degree}25'27''.06$ (J2000). Towards the North, a third local maximum is visible, with a peak value of $\log[$N(NH$_3$)]$_{peak} = (14.26 \pm 0.15)$ at position R.A.=$17^h11^m23^s.00$, Dec. = $-27^{\degree}24'40''.6$ (J2000). The mean value across the whole source is $\langle \log[ \text{N(NH}_3) ]\rangle= 14.23 $.

\par

\begin{figure} [!h]
   \centering
   \includegraphics[width= \hsize]{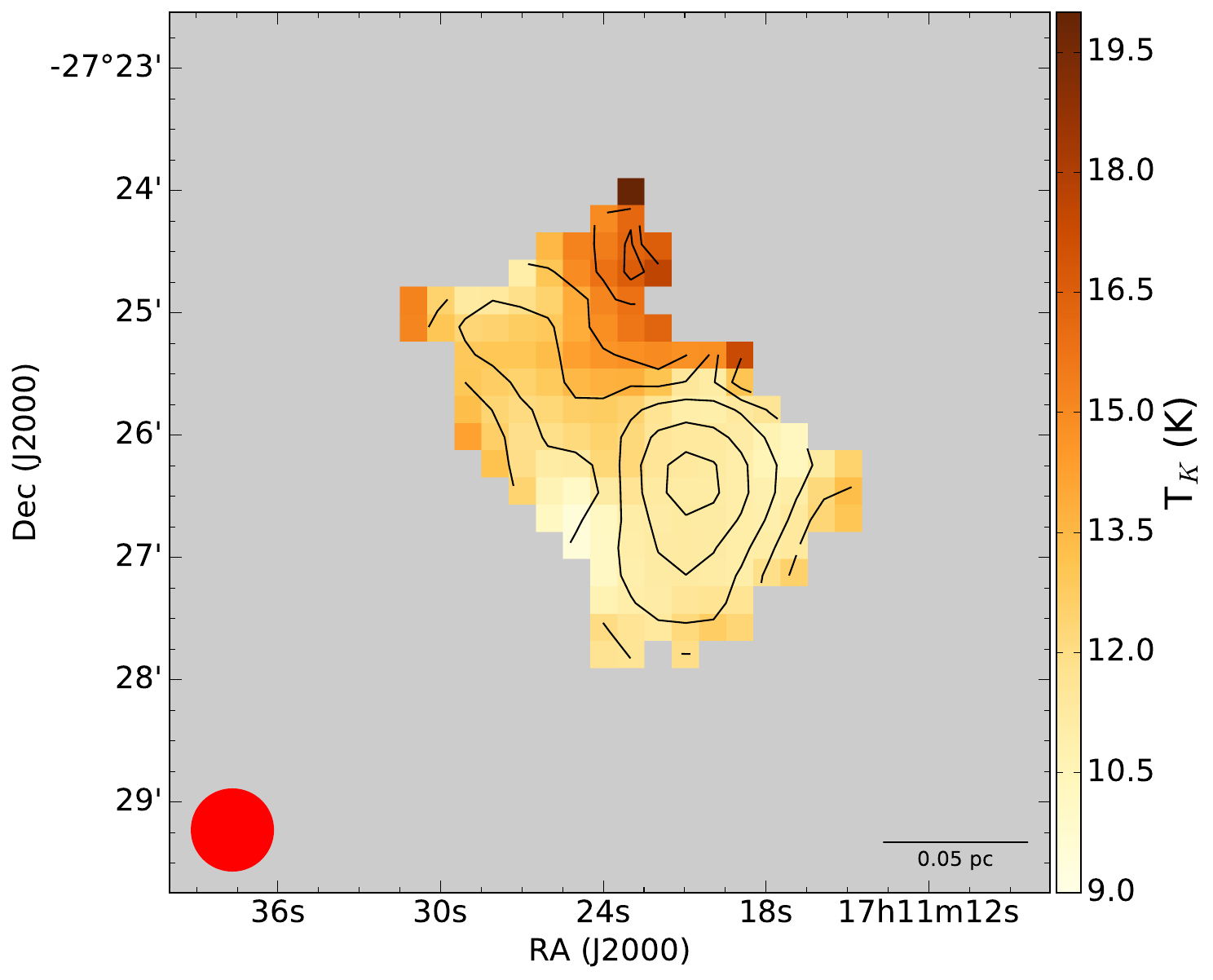}
   \caption{\tk map of the source. The parameter median uncertainty is $0.5\,$K. Contours show the \am column density $\log[$N(NH$_3$)/cm$^{-2}$]   at the following levels: $[14.0, 14.2,14.4,14.6,14.75]$.}
  \label{tkin}
   \end{figure}

Figure \ref{tkin} shows the kinetic temperature map. The densest part of the cloud is also the coldest: $\langle {\text{T}}_K \rangle= 11.3  \,$K for regions where $\log[$N(NH$_3)]>14.4$, a typical value for cold cores heated only by the external radiation field and cosmic rays \citep{Zucconi01,Evans01}. In the northern part of the cloud, temperatures are higher, reaching $16-17\,$K, which possibly indicates another source of heating in addition to the standard cosmic ray field or surrounding radiation field.
   \par
\begin{figure} [!h]
   \centering
   \includegraphics[width= \hsize]{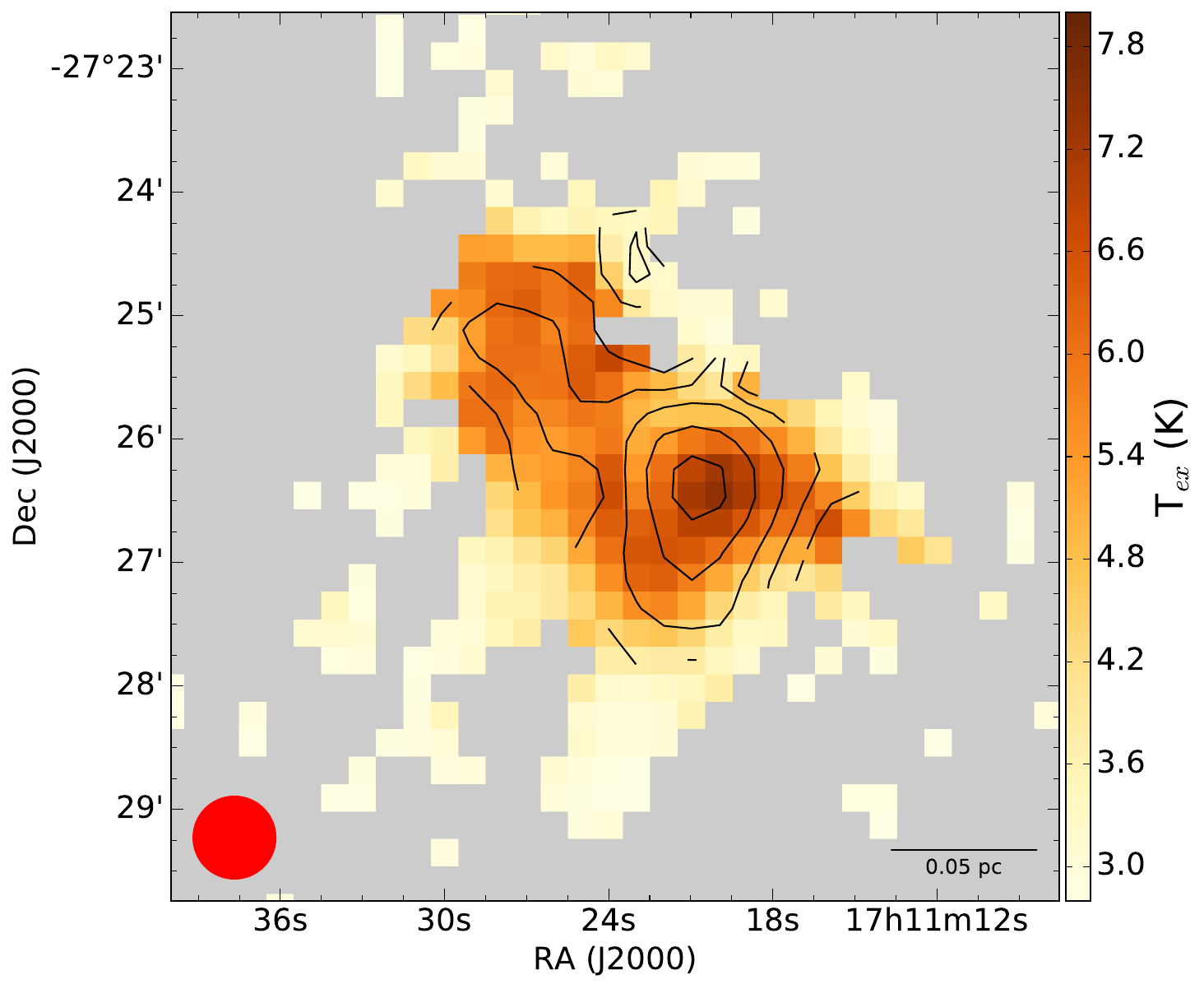}
   \caption{\tex map in B59. The median uncertainty is $0.3\,$K. The contours are the same as in Figure \ref{tkin}.}
  \label{tex}
   \end{figure}

 The excitation temperature, which regulates the population levels within each $K$-ladder, is shown in Figure \ref{tex}. The data are more spatially extended than the previous two maps due to the flagging criteria on integrated intensity SNR (i.e. the (1,1) transition has more pixels meeting these criteria than the (2,2)). High column density regions show higher  \tex since the upper state becomes more populated through collisions.   The mean value across the entire map is  $\langle \text{T}_{ex} \rangle = 4.83  \,$K, but it rises to $\langle{\text{T}}_{ex} \rangle= 5.99 \,$K  at regions where $\log[$N(NH$_3)]>14.4$. The peak value is   T$_{ex}^{peak} = (7.46 \pm 0.05) \,$K, found at position R.A.=$17^h11^m19^s.99$, Dec. = $-27^{\degree}26'27''.3$ (J2000), within a beam width from the column density peak. It is interesting to notice that generally \tk is $\approx 7 \,$K greater  T$_{ex}$, which suggests that the molecule is subthermally excited (i.e. the volume density averaged within the beam is lower than the critical density). 
   \par
\begin{figure} [!h]
   \centering
   \includegraphics[width=  \hsize]{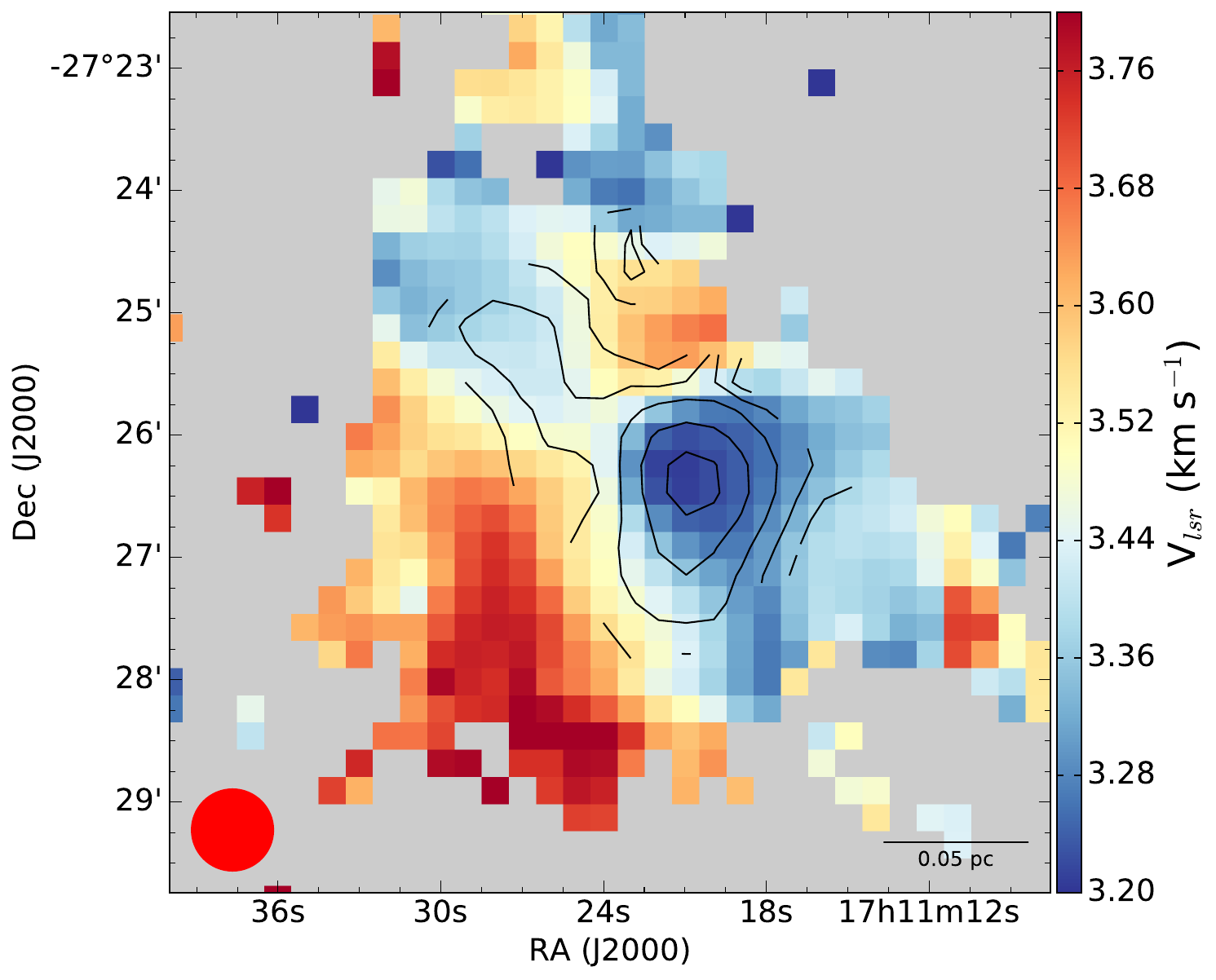}
   \caption{\vc map in B59. The median uncertainty across the map is $0.03\,$km$\,$s$^{-1}$.  Contours show \col levels as in Figure \ref{tkin}.}
  \label{Vlsr}
   \end{figure}
\begin{figure} [!h]
   \centering
   \includegraphics[width=  \hsize]{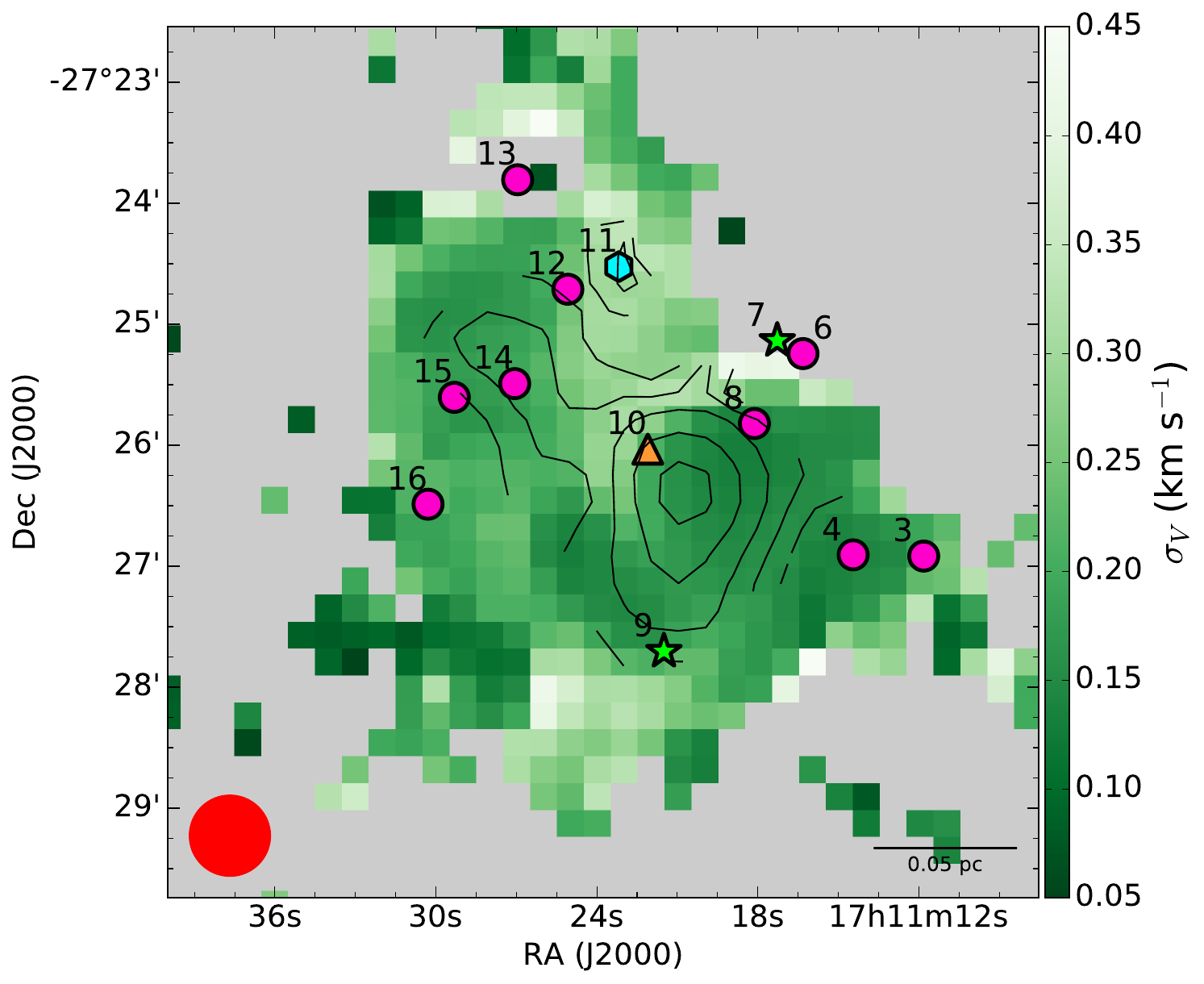}
   \caption{Intrinsic velocity dispersion, \sig, map of B59 derived from the line fit. The median uncertainty is $0.03\,$km$\,$s$^{-1}$. Contours show \col levels as in Figure \ref{tkin}. The markers show the YSOs positions labeled following \cite{Brooke07} classification (Cyan hexagon: Class 0/I. Orange triangles: Class I. Green stars: Class I/II. Magenta circles: Class II).}
  \label{sig}
   \end{figure}

 The centroid velocity and velocity dispersion maps are shown in Figure \ref{Vlsr} and Figure \ref{sig}, respectively. \vc has values in the range $\approx [3.0 - 3.8]\,$km$\,$s$^{-1}$, in good agreement with previous observations that report an ambient velocity of $3.6\,$\kms \citep{Onishi99,DuarteCabral12}. The densest and coldest part of the core shows a kinematically coherent structure, with a mean value $\langle{V_{lsr}} \rangle= (3.29 \pm 0.07) \,$\kms (computed on pixels with $\log[$N(NH$_3)]>14.4\,$). \par 
The velocity dispersion range is $\approx [0.05-0.50]\,$km$\,$s$^{-1}$. The mean value is $\langle{\sigma_V} \rangle= 0.21 \,$km$\,$s$^{-1}$, and it decreases to $\langle{\sigma_V} \rangle= 0.18 \,$\kms in the densest part of the core.   
 
 \newpage

 \section{Protostellar feedback \label{ProtoFeed}}
 \cite{Brooke07} studied the B59 region using Spitzer data belonging to the c2d survey, and found approximately 20 young stellar objects possibly associated with B59. The authors derived each source's spectral class according to the SED slope, following the prescriptions introduced by \cite{Lada84} and developed further by \cite{Andre00}. The latter is defined as $\alpha = d \log(\lambda F_{\lambda})/d \log(\lambda)$ between $2.2\, \mu$m and $8.0\, \mu$m. This parameter traces the evolutionary stage of YSOs, since as time passes by the contribution from the central object becomes more important than the dusty envelope emission at larger wavelengths. Hence, the SED slope changes from positive to negative. Class I sources are identified by $\alpha >0.3$, while Class II are represented by $\alpha < -0.3$. Sources with $-0.3<\alpha <0.3$ are classified as flat (or I/II). Finally, Class 0 objects present $T_{bol} <70\,$K and $L_{smm}/L_{bol}>0.5\%$, i.e. a large fraction of their emission is at submillimeter wavelengths.
 \par
  We refer to the Table 2 of \cite{Brooke07} for the sources' positions, identification names, and spectral classes (after the correction for extinction, when available). To analyze the impact of protostellar feedback from the YSOs into the ambient cloud, we search for a correlation between velocity dispersion and the YSOs positions. As one can see in Figure \ref{sig}, evolved Class II sources are mostly located at lines of sights of low velocity dispersion. This correspondence suggests that these objects are possibly no longer embedded, being either foreground or background sources. On the other hand, some of the less evolved objects are located in zones of broader lines or at the edge of a transition area. In particular, BHB07-11 (B11, labeled as \#11 in the Figures), one the youngest sources in the protocluster \citep{Hara13}, is found in one of the zones with the highest \sig values. It is useful to compare these values with the isothermal sound speed, which at T$_K=11\,$K, approximately the average kinetic temperature at the high density regions, is:
\begin{equation}
\label{CS}
C_S = \sqrt{\frac{k_B \text{T}_K}{\mu m_{\text{H}}}} = 0.20 \, \text{\kms},
\end{equation}
where $\mu = 2.33$ is the mean molecular weight of the gas (which is assumed to be composed mostly by H$_2$ and a fraction of 10\% of He), $k_B$ is the Boltzmann constant, and $m_{\text{H}}$ is the mass of the hydrogen atom. The densest part of B59 is then characterized by marginally subsonic velocity dispersion ($\langle{\sigma_V} \rangle= 0.18 \,$km$\,$s$^{-1}$, see Sec. \ref{ParaMaps}), and appears to be a quiescent and coherent structure. We will thus refer to it as a \textit{coherent core} \citep{Goodman98}.
   \par
\begin{figure} [!h]
   \centering
   \includegraphics[width=   \hsize]{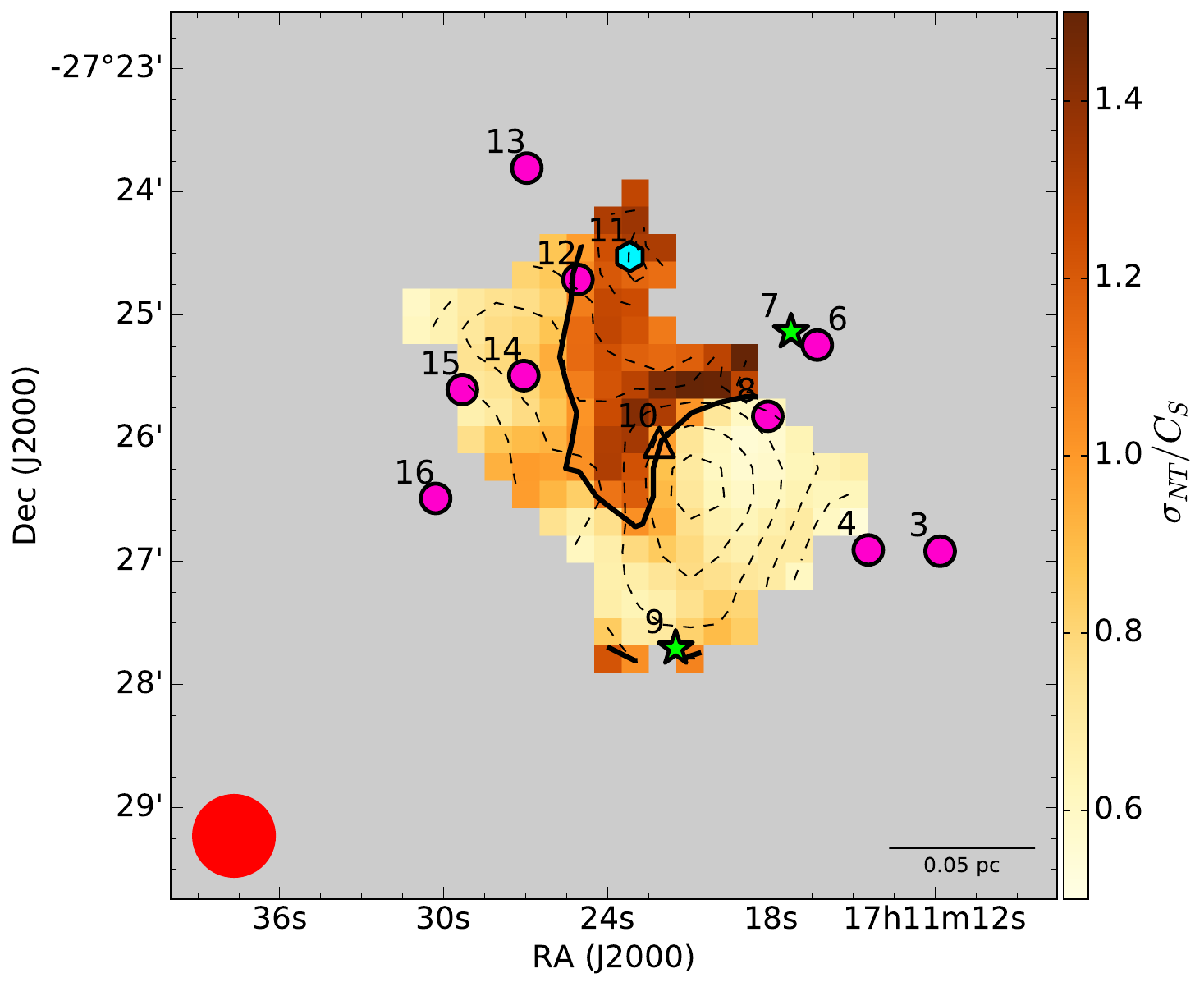}
   \caption{The obtained map of $\sigma_{NT}/C_S$, which has a median uncertainty of $0.04$. The black line shows the transition from subsonic to supersonic regime ($\sigma_{NT}/C_S = 1.0$). The YSO positions and the \am column density contours are indicated as in the previous image.}
  \label{TurbRatio}
   \end{figure}

In order to quantify the amount of turbulence in the source, we separate the thermal $\sigma_{Th}$ and non-thermal $\sigma_{NT}$ components of the velocity dispersion \citep{Myers91}. The former is calculated through:
\begin{equation}
\sigma_{Th}  = \sqrt{\frac{k_B \text{T}_K}{\mu_{\text{NH}_3} m_{\text{H}}}}  \; , 
\end{equation}
where $\mu_{\text{\am}}$ is the ammonia molecular weight (equal to 17.03) Assuming that the two quantities are independent, they add in quadrature,
\begin{equation}
\sigma_V^2 = \sigma_{Th}^2 + \sigma_{NT}^2  \quad \rightarrow \quad  \sigma_{NT} = \sqrt{\sigma_V^2 - \sigma_{Th}^2} \; .
\end{equation}
$\sigma_{NT}$ takes into account all gas motions unrelated to thermal effects, such as turbulence. Dense cores are argued to form from large-scale ($> 0.1 \,$pc) turbulence dissipation due to shock compressions \citep{Elmegreen04, MacLow04}. It is therefore important to spatially resolve the internal turbulence in the core, and to try to correlate it with the effects of protostellar feedback. We thus compute the $\sigma_{NT}/C_S$ ratio pixel by pixel, making use of Equation \ref{CS} and the kinetic temperature derived from our molecular line fitting and show the results in Figure \ref{TurbRatio}.
A large portion of the coherent core is characterized by subsonic motions: the ratio reaches a minimum value of $ \sigma_{NT}/C_S =0.54 \pm 0.05$, and the mean for pixels with $\log[$\col$]>14.4\,$ is $ \langle \sigma_{NT}/C_S  \rangle =0.80$. It is interesting to note that, within our map coverage, only class II sources are found in the line-of-sight of subsonic motions, reinforcing the idea that they are off-core objects. On the other hand, younger objects are located on the edge of the transonic transition (B09 or B10) or in zones of supersonic motions (B11).  The transition between subsonic and supersonic media appears to be sharp, as found by other studies \citep{Foster09,Pineda10}. The maximum value found for the $\sigma_{NT}/C_S$ ratio is $1.65 \pm 0.31$ . The enhancement of non-thermal motions in the North is most likely due to the B11 (and possibly B09 and B10) bipolar outflow, which is injecting turbulence in the core as also seen with hydrodynamic simulations \citep{Offner14}. In fact, one of the outflow lobes is spatially correlated with the region with the highest $ \sigma_{NT}/C_S $ values (see Fig. 8 from \citealt{DuarteCabral12}).
   \par

\begin{figure} [!h]
   \centering
   \includegraphics[width= \hsize]{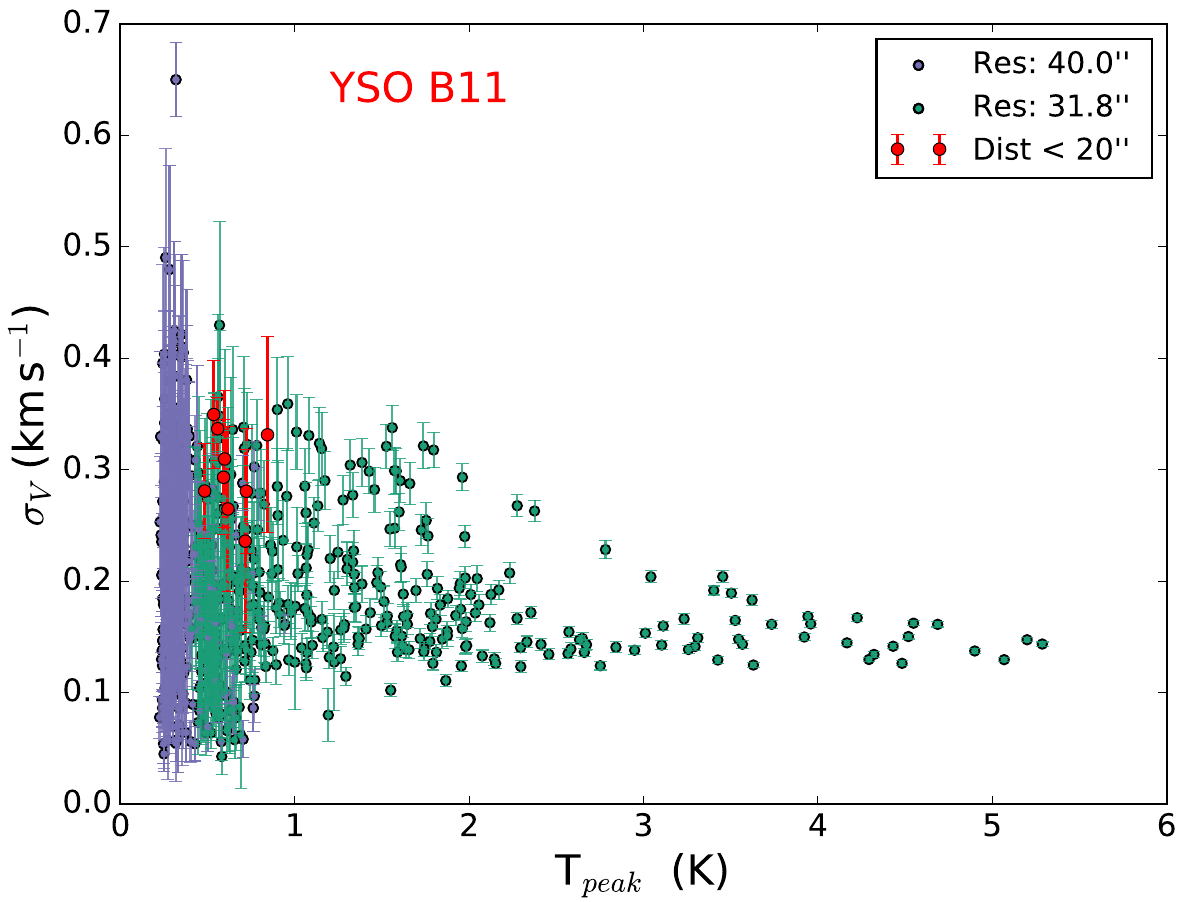}
   \caption{Scatter plot of \sig as a function of T$_{peak}$. Points from the original, higher resolution maps are shown with green dots, whilst violet ones come from the smoothed and regridded maps and are used when the former are not available. Red dots are positions =within a beam size from B11.}
  \label{ScatSigB11}
   \end{figure}

In order to inspect the correlation between \sig or $\sigma_{NT}/C_S$ and the YSOs positions, we have produced scatter plots of these two quantities and T$_{peak}$ (the peak temperature in the (1,1) spectra). An example is shown in Figure \ref{ScatSigB11}, which focuses on B11. We have used two different data sets: where available, we have taken points from the original (unsmoothed and unregridded, native resolution $=31.8''$) map (green dots) while for the outskirts of the source, where the original SNR is low, we have used the smoothed and regridded maps (violet dots), which have a resolution of $40''$ (see Sec. \ref{Smoo}). The red points indicate positions within a smoothed beam size from the YSO B11 (i.e. the distance from the latter is less than $20''$). A general trend is seen: at high T$_{peak}$ (which correlates with the \am column density), the velocity dispersion is often below the sonic speed ($\approx 0.20\,$km$\,$s$^{-1}$). For T$_{peak} < 2\,$K, \sig values increase, in a similar way as reported by \cite{Pineda10}. The plot reveals also a trend of the velocity dispersion in the YSO surroundings, as suggested by the fact that red points are mainly located at high velocity dispersions. We have performed the same analysis on all the sources covered by our data, computing the mean \sig and dispersion over a beam size around each one. Results are summarized in Table \ref{YSOfeed}. \par

\begin{figure} [!h]
   \centering
   \includegraphics[width= \hsize]{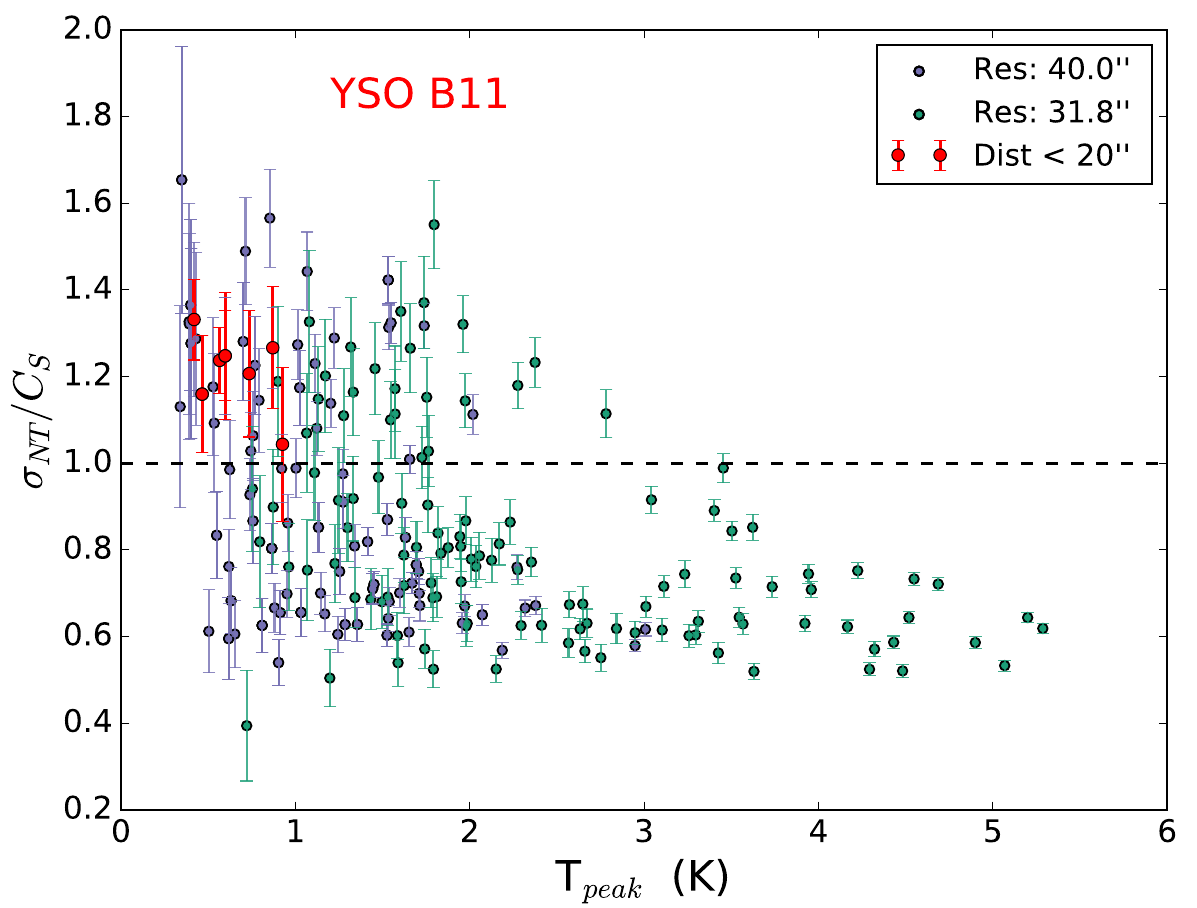}
   \caption{Scatter plot of the ratio between $\sigma_{NT}$ and $C_S$ versus T$_{peak}$. Color code is the same as in Figure \ref{ScatSigB11}. The black, dashed line marks the transonic level.}
  \label{ScatTurbB11}
   \end{figure}

We have also studied how the $\sigma_{NT}/C_S$ ratio correlates with T$_{peak}$, and Figure \ref{ScatTurbB11} shows the respective scatter plot. The quiescent nature of the coherent core is clearly visible at high T$_{peak}$. Red points, as previously described, indicate pixels closer than $20''$ to B11. We have computed the mean and standard deviation of the ratio in the surroundings of all the available YSOs. Table \ref{YSOfeed} summarizes these values as well. \par
Class II objects usually present low mean velocity dispersion values if compared with the sonic speed, and, where available, their mean $ \sigma_{NT}/C_S $ is less than 1.0.There are two exceptions: B03, which is located very close to the edge of our map coverage, and B12, which though is too close to B11 to be disentangled with our smoothed angular resolution. Among the younger sources, on the other hand, B09, B10, and B11 are found in a more turbulent environment. B10 and B11, in particular, show the highest values for both the analyzed variables. This can suggest that they are truly embedded in the cloud and that they are affecting through protostellar feedback the surrounding gas.

\begin{deluxetable}{cccc}
\tabletypesize{\footnotesize}
\tablecaption{Results from the analysis of \sig and $\sigma_{NT}/C_S$ ratio maps in correlation with YSOs positions. Missing values represent sources not covered by the \tk map. The YSOs spectral classes from \cite{Brooke07} are also indicated. \label{YSOfeed}}
\tablecolumns{4}
\tablehead{ \colhead{ YSO id} & \colhead{IR spectral class }  & \colhead{\sig (\kms) }  & \colhead{ $\sigma_{NT}/C_S$}}
\tablewidth{0pt}
\startdata
B03 & II 				&$0.21 \pm 0.06$& - \\ 
B04 & II 				&$0.14 \pm 0.01$ &  - \\
B08 &II	&$0.16 \pm0.05$ & $0.58\pm0.07$\\
B09 & I/II 				&$0.22 \pm0.05$  &$0.84\pm0.18$\\
B10& I	&$0.23 \pm 0.06$ & $1.06 \pm 0.29$ \\
B11 & 0/I 				&$0.30 \pm 0.03$ &$1.21 \pm 0.08$\\
B12 & II	&$0.26 \pm 0.06$ &$0.94 \pm 0.17$ \\
B14 & II 				&$0.19 \pm0.03 $&  $0.80 \pm 0.12$\\
B15 & II 				& $0.17\pm 0.02$ & $0.71\pm 0.06$\\
B16 & II 				& $0.18 \pm0.04$ & - \\ 
 \enddata
 \end{deluxetable}

\newpage
   
\section{Comparison between H$_2$ and \am \label{CompHersch}}
\subsection{Column density and temperature \label{H2Comp}}
 \am (1,1) and (2,2) transitions are usually a tracer of cold ($10-20\,$K) gas (\citealt{HoTownes83}), as shown by many observational works  \citep{Rosolowsky08, Friesen09, Pineda15,  FriesenPineda17}. Using temperature as a proxy to assess the coupling between molecular gas and dust, we accessed Herschel SPIRE data to determine maps of H$_2$ column density and dust temperature \tdust, following the approach described in Appendix \ref{Herschel}. The SPIRE maps at $250\, \mu$m and $350\, \mu$m were smoothed to the resolution of the $500 \, \mu$m one, leading to a common angular resolution of FWHM$=38.4 ''$, which is comparable to the resolution of the smoothed GBT data. Furthermore, we performed a simple background subtraction by computing the average flux density in the surrounding of the source (d$ \; \lesssim 12' $) for each wavelength.  The $3 \, \sigma$ contour ($\approx0.18\,$K$\,$km$\,$s$^{-1}$) in the (1,1) integrated intensity map was chosen to set the external edge of the source. After a few tests, this region selection to compute the background contribution appeared to be the best to eliminate the warmer surrounding medium component, which otherwise would affect the estimation of the column density and dust temperature\footnote{We also tested a background subtraction approach consisting in a cut of low spatial frequency components in the Fourier space. This gave results consistent to the ones obtained with the method described in the text, within the uncertainties.}. Finally, the maps were regridded to the same pixel size of the ammonia ones, to allow one-to-one comparisons of the two datasets.\par
\begin{figure} [!h]
   \centering
   \includegraphics[width= \hsize]{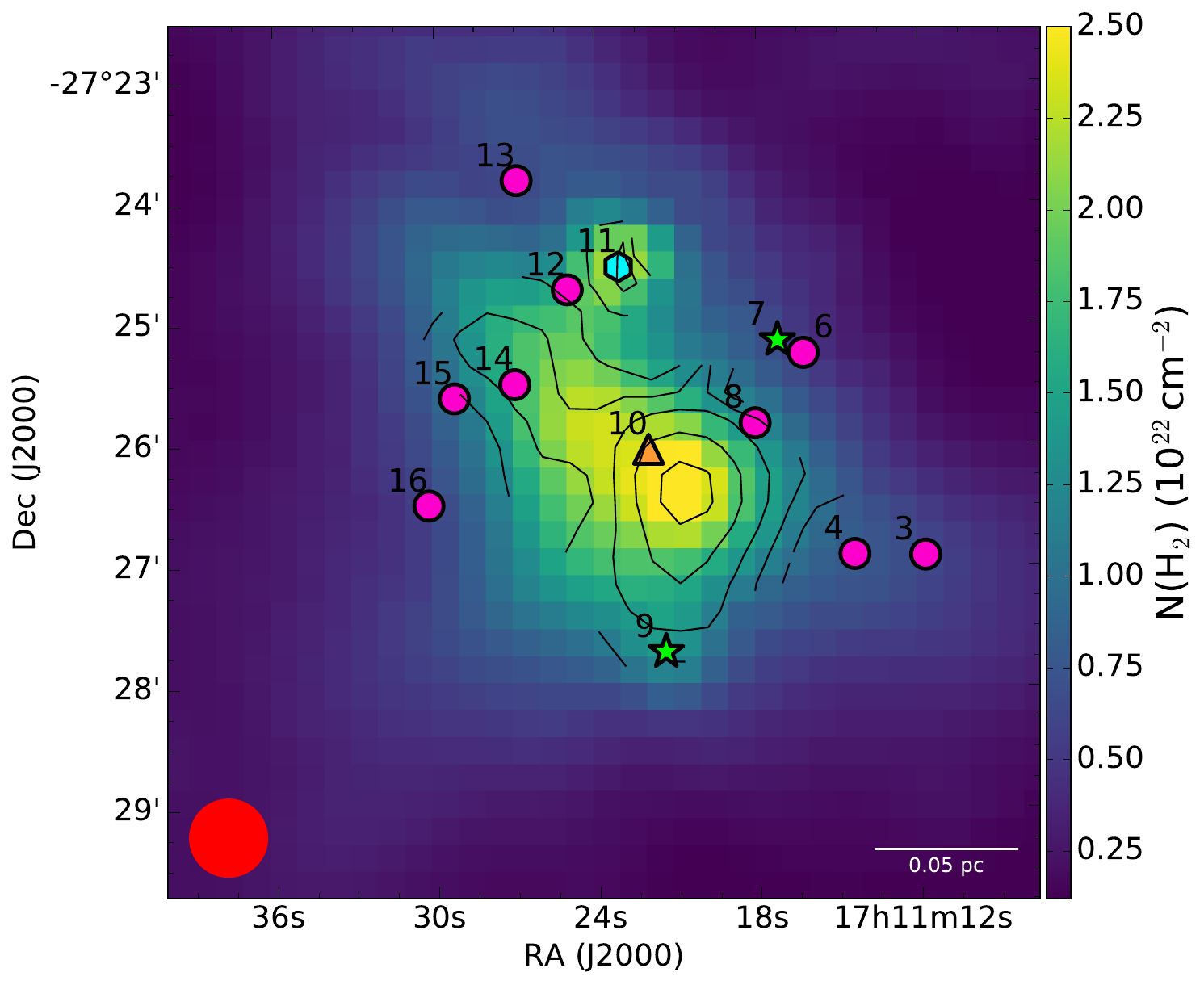}
   \caption{H$_2$ column density map. The median uncertainty is $0.04 \cdot 10^{22}\,$cm$^{-2}$. Contours represent $\log[$N(NH$_3$)] as in Figure \ref{tkin}. YSO positions are also indicated, as defined in Figure \ref{sig}.}
  \label{NH2}
   \end{figure}

 Figure \ref{NH2} shows the derived N(H$_2$) with contours from \col. As expected, the two quantities present a similar distribution.  The N(H$_2$) peak at $\log[\text{N(H}_2)] = (22.43\pm0.06)$ is found within a beam size of the \col peak. The average value computed in the coherent core ($\log[$N(NH$_3)]>14.4$) is $\langle  \log[\text{N(H}_2)]\rangle  = (22.30 \pm 0.09)$. 
Both maps do not show a clear correlation with the young sources' positions, with the exception of B11, which is visible as an enhancement in both the \am and H$_2$ data. If we take into account a circular region of $25''$ of radius around that YSO, we find mean values of $\langle  \log[\text{N(H}_2)]\rangle  = (22.26\pm 0.07)$ and $\langle \log[ \text{\col}] \rangle  = 14.04 $. In the main part of the core, this level of H$_2$ column density corresponds to ammonia values of at least $\log  [\text{N(NH}_3)]\approx 14.2$. This difference can suggest that \am is depleted in B11, possibly due to chemical effects. For example at \tdust$\approx20\,$K, CO molecules, which have previously suffered from freeze-out onto dust grains, start to evaporate and return to the gas phase, destroying the \am precursors \citep{CharnleyRodgers02, RodgersCharnley08}. \par
\begin{figure} [!h]
   \centering
   \includegraphics[width=  .9\hsize]{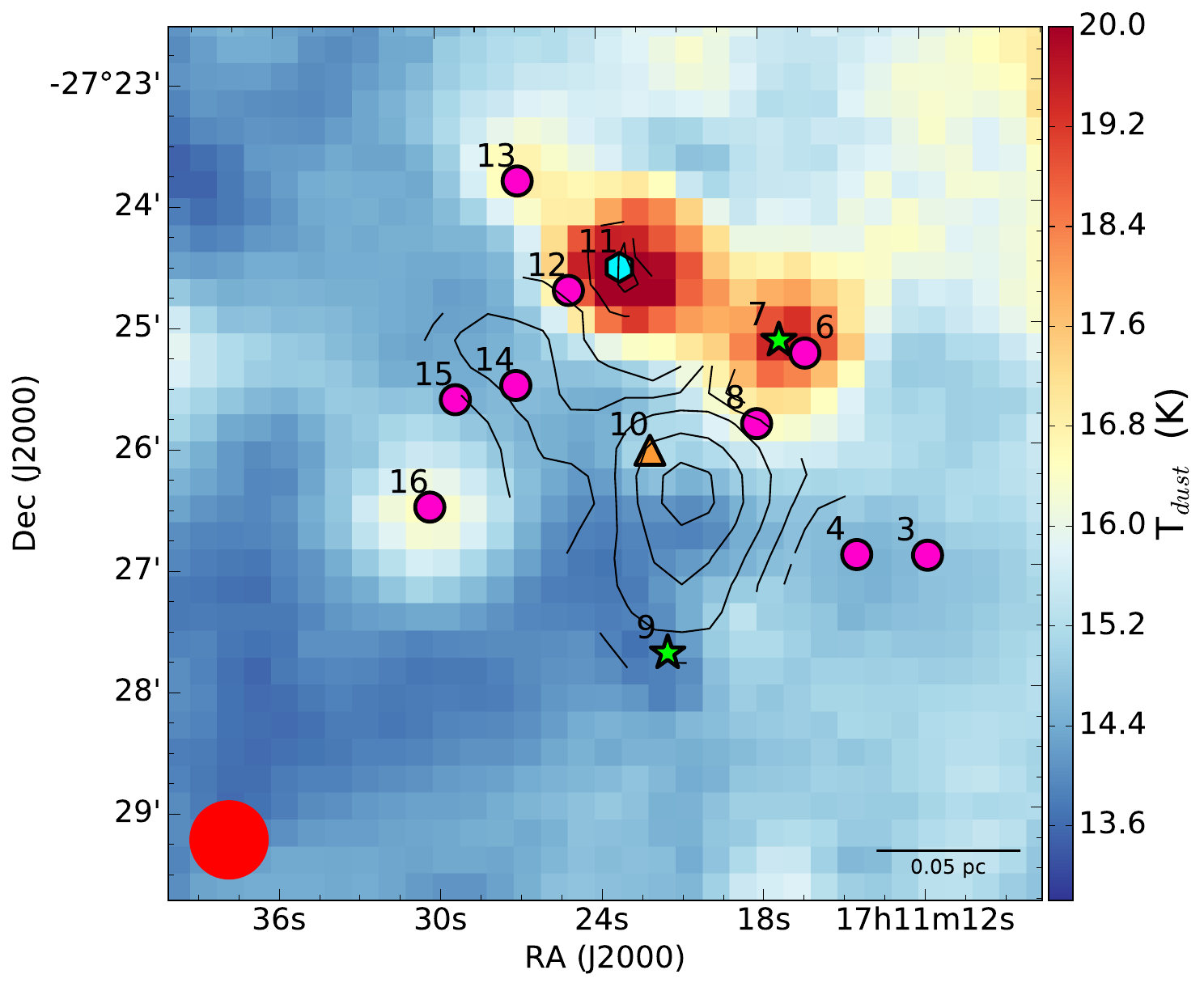}
   \caption{Herschel based \tdust map with overlaying YSOs positions and \col contours, as in the previous Figures. The characteristic uncertainty is $0.7$\,K.}
  \label{Tdust}
   \end{figure}

The CO evaporation is possibly due to internal heating, as suggested by the \tdust  map shown in Figure \ref{Tdust}. The image shows few warmer condensations in proximity of some YSOs. B11 is the hottest spot, with a peak value of \tdust$= (21.8 \pm 2.2)\, $K, and an average (within a beam size) of $\langle \text{T}_{dust} \rangle = (20.6 \pm 1.1)\, $K. At this temperature, CO starts to evaporate from dust and impact the ammonia abundance \citep{Tielens05}. For the surroundings of B06 and B07 (which cannot be disentangled with our spatial resolution), we find$\langle \text{T}_{dust} \rangle = (18.8\pm0.7) \, $K. Around B13 and B16 we obtain $\langle\text{T}_{dust} \rangle = (16.5\pm0.4) \, $K and $\langle \text{T}_{dust} \rangle = (16.2\pm0.2) \, $K, respectively. None of these sources, however, is covered by the \am data with the exception of B11. The mean temperature observed for the entire source, using a threshold of $\log[\text{N(H}_2)] >21.6$, is $\langle\text{T}_{dust} \rangle =  (15.2\pm1.4) \, $K.   
\par
\begin{figure} [!h]
   \centering
   \includegraphics[width=  .9\hsize]{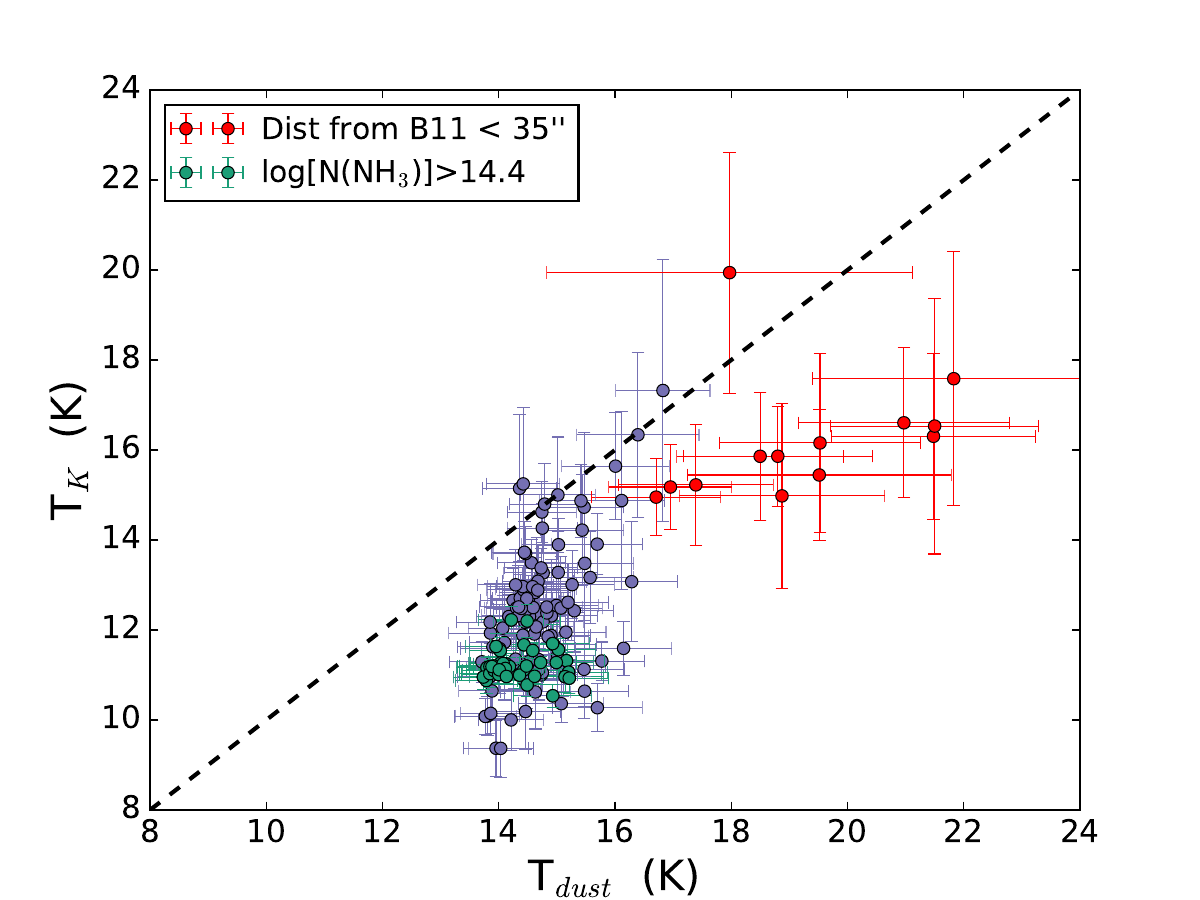}
   \caption{Scatter plot of \tk versus \tdust. The black, dashed line represents the 1:1 relation. Green points indicate the coherent core ($\log[\text{\col}] >14.4$), red ones are closer than $35''$ from B11 and violet dots come from the rest of the source. }
  \label{TkTdust}
   \end{figure}

Figure \ref{TkTdust} shows a scatter plot of \tk and \tdust. 
The dashed curve denotes the line of unity, and it is interesting that almost all points lie below it within errorbars. This difference means that dust temperature overestimates the kinetic one, likely because Herschel is more sensitive to more extended, warmer emission than ammonia observations (even after excluding background and foreground emission). As a result, the dust temperature, which is an average along the line of sight, is systematically higher than the kinetic temperature traced by \am, which traces mostly the dense portions of B59. Green dots indicate the coherent core, and they confirm the presence of an almost isothermal structure. Red dots show positions closer than $35''$ around B11 (a circular region that comprehend most of the hot, clumpy feature around the YSO visible in Figure \ref{Tdust}), and as expected they present the highest values, supporting the scenario where this source is actively heating its surroundings, also because \tdust decreases away from B11 center in all directions, thus excluding possible enhancements due to the external irradiation from the nearby Sco OB2 association.
   
\subsection{Ammonia abundance}
\begin{figure} [!h]
   \centering
   \includegraphics[width=  \hsize]{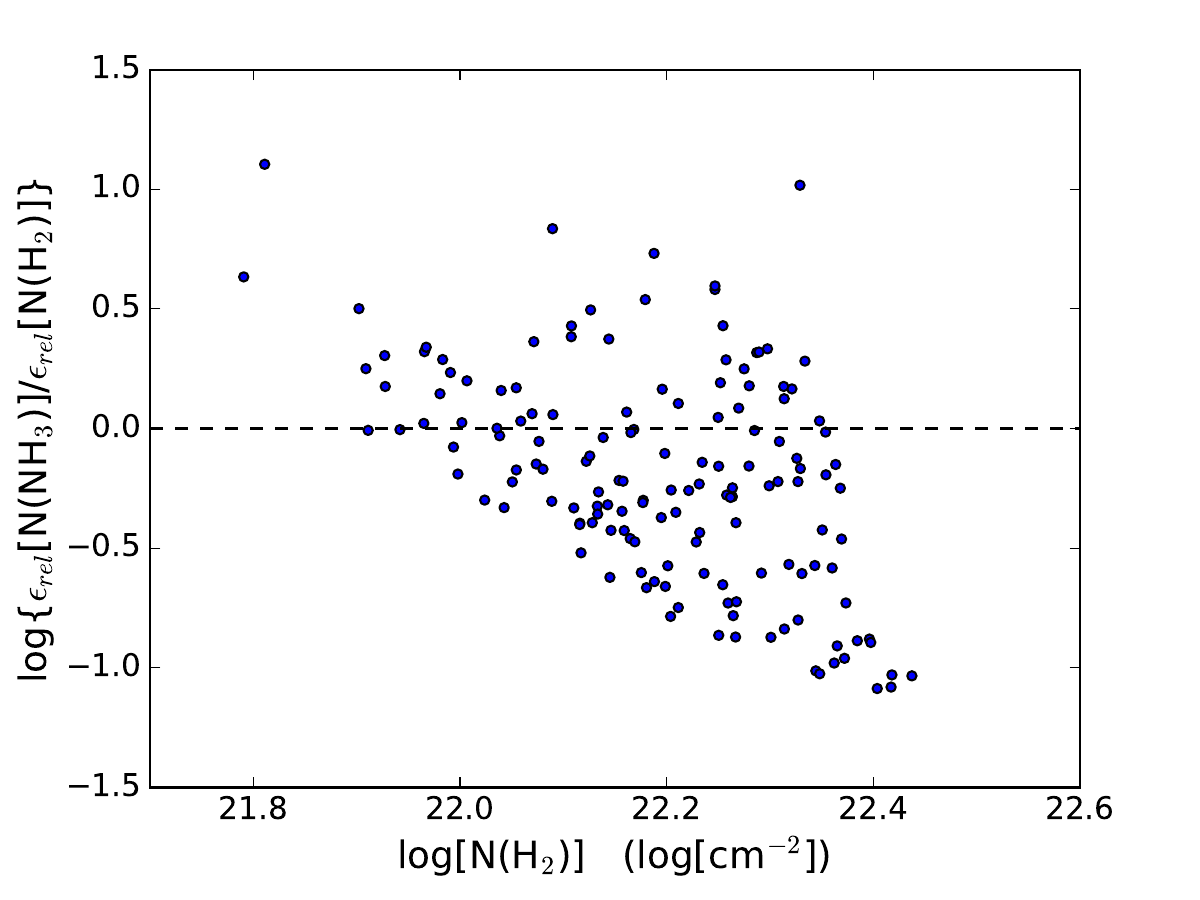}
   \caption{Ratio of the  \col relative uncertainties ($\epsilon$N(NH$_3)$/N(NH$_3)$) over the N(H$_2$) ones in logarithmic scale, plotted versus log[N(H$_2$)]. The black dashed line indicates where the ratio is equal to 1.0, below which the majority of the points is found.  }
  \label{CompErr}
   \end{figure}

The canonical way to compute ammonia molecular abundance consists of taking the ratio between its column density and the H$_2$ column density: \xam = \col / N(H$_2$). A more robust approach involves the so called zero point column density, i.e., the minimum H$_2$ column density at which \am starts to be detected, indicated as N(H$_2)_0$ (see \citealt{Pineda08} for a similar approach for CO isotopologues). Therefore, the relation between \am and H$_2$ column density in linear scale reads:
\begin{equation}
\label{cols}
\text{\col}= \text{X(NH}_3) \cdot \text{N(H}_2) + K
\end{equation}
where $K$ is a negative constant related to N(H$_2)_0$. We perform a linear least squares regression fit to the data to obtain both $K$ and the molecular abundance. For convenience, we inverted Equation \ref{cols} and fitted the relation $  \text{N(H}_2)  = m \cdot \text{\col} + \text{N(H}_2)_0$ (where $m = 1 / \text{\xam}$) to consider only uncertainties on N(H$_2$), which are usually larger that those on \col. Figure \ref{CompErr} shows the ratio between the relative uncertainties on \col ($\epsilon_{rel}$[N(NH$_3$)]) and the ones on N[H$_2$] ($\epsilon_{rel}$[N(H$_2$)])\footnote{The relative uncertainty $\epsilon$ of a quantity ($X\pm \sigma_X$) is computed as $\epsilon[X] = \sigma_X / X$.}. The results of the linear fit are listed in Table \ref{fitRes}.

\begin{figure} [!h]
   \centering
   \includegraphics[width=  \hsize]{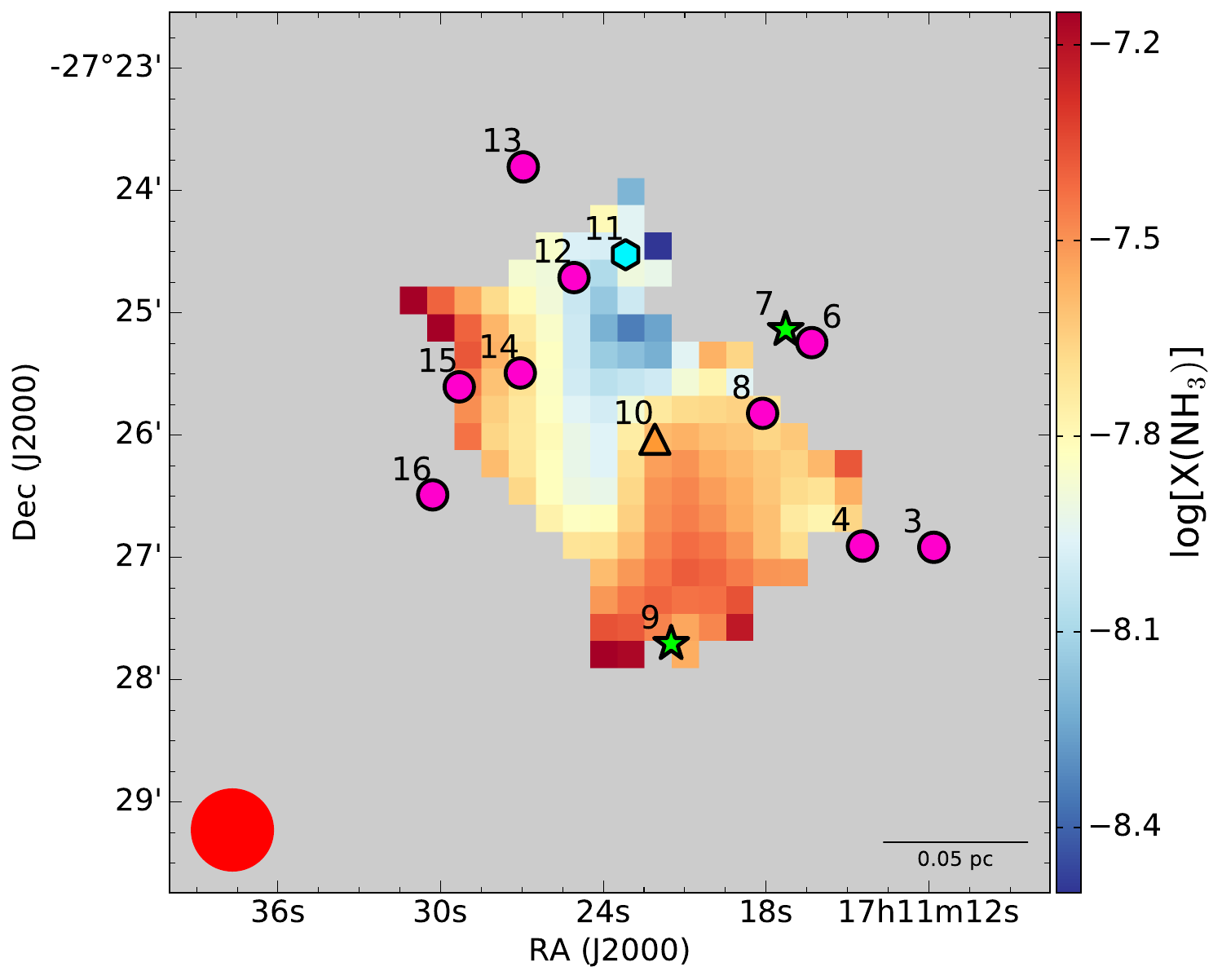}
   \caption{\xam map obtained with the approach described in text. The median uncertainty is $0.3 \cdot 10^{-8}$. YSOs positions are also indicated. }
  \label{AmAbund}
   \end{figure}

\begin{deluxetable}{cc}[h]
\tablecaption{Best fit parameters for N(H$_2$)-\col relation\label{fitRes}}
\tablecolumns{2}
\tablehead{ \colhead{Parameter} & \colhead{Value}}
\setlength{\tabcolsep}{16pt}
\startdata
$m$  ($10^7$) & $(3.45 \pm 0.26)$  \\
 N(H$_2)_0$ ($10^{21}$cm$^{-2}$)    &   $(6.38 \pm 0.38)$  \\
\xam  ($10^{-8}$)& $ (2.90 \pm 0.22)$\\
$A_V$\tablenotemark{a} (mag) & $(6.79 \pm 0.40)$  \\
$r$\tablenotemark{b}   &  0.619  \\
\enddata
\tablenotetext{a}{Extinction in magnitude calculated using the relation
N(H$_2)=9.4\,10^{20}  \left( A_V/\text{mag} \right) \text{cm}^{-2}$ \citep{Bohlin78}.} 
\tablenotetext{b}{Linear correlation coefficient.}
\end{deluxetable}
The zero point column density found is in agreement with the value of $A_V = 7\,$mag used by the GAS survey as a threshold value for detection of \am emission. The \xam value we obtain has to be considered an average across the source, and it is in good agreement with other low-mass star forming regions, such as Ophiuchus (\xam$=1 - 10 \cdot 10^{-8}$, \citealt{Friesen09}) and Perseus (\xam$=1- 8 \cdot 10^{-8}$, \citealt{Foster09}). With the parameters just obtained, we can also recover the abundance map, computing pixel by pixel:
\begin{equation}
\text{\xam} = \frac{\text{\col}}{   \text{N(H}_2) -\text{N(H}_2)_0 }
\end{equation}
The spatial distribution of \xam is shown in Figure \ref{AmAbund}.  Interestingly, the ammonia abundance near the position of B11 is very low compared to the densest coherent core, confirming the \am depletion in this area mentioned in the previous sections.

\section{Core dynamical state \label{VirialAna}}
We investigate the core dynamical state by means of the virial ratio, usually defined as $\alpha_{vir} = M_{vir} /M $ \citep{Bertoldi92}, where $M$ and $M_{vir}$ are the source mass and virial mass, respectively. If we consider only the contribution of the kinetic energy $K$ and the gravitational potential energy $U$ to the source stability, the last equation can be expressed as:
\begin{equation}
\alpha_{vir} = \frac{2K}{|U|}  \; .
\end{equation}
The kinetic term, which comprises both thermal and turbulent motions, is computed as:
\begin{equation}
K = \frac{3}{2} M \sigma_{vir} ^2 \; ,
\end{equation}
where M is the mass and $\sigma_{vir} $ the total, one-dimensional velocity dispersion of the gas (the factor $3$ takes into account components on the other two spatial directions, assuming an isotropic velocity distribution). $\sigma_{vir}$ can be derived assuming that the non-thermal component of the overall molecular gas velocity is the same as in the \am one. Then:
\begin{eqnarray}
\sigma_{vir}^2 &  =& \sigma_{\text{tot}, \text{NT}} ^2 + \sigma_{\text{tot}, \text{T}} ^2  \; , \nonumber \\ 
\sigma_{vir} ^2 & =   & \sigma_V^2 -  \sigma_{\text{NH}_3, \text{T}}^2 + \sigma_{\text{tot}, \text{T}}^2   \;  =  
 \sigma_V^2 -\frac{k_B \text{T}_K}{\mu_{\text{NH}_3} m_{\text{H}}} + \frac{k_B \text{T}_K}{\mu m_{\text{H}}} \; .
\end{eqnarray}
The velocity dispersion is computed pixel by pixel, and the mean value over the obtained map is $\sigma_{vir}= (0.29 \pm 0.05)\,$km$\,$s$^{-1}$.  \par
\begin{figure} [!h]
   \centering
   \includegraphics[width= \hsize]{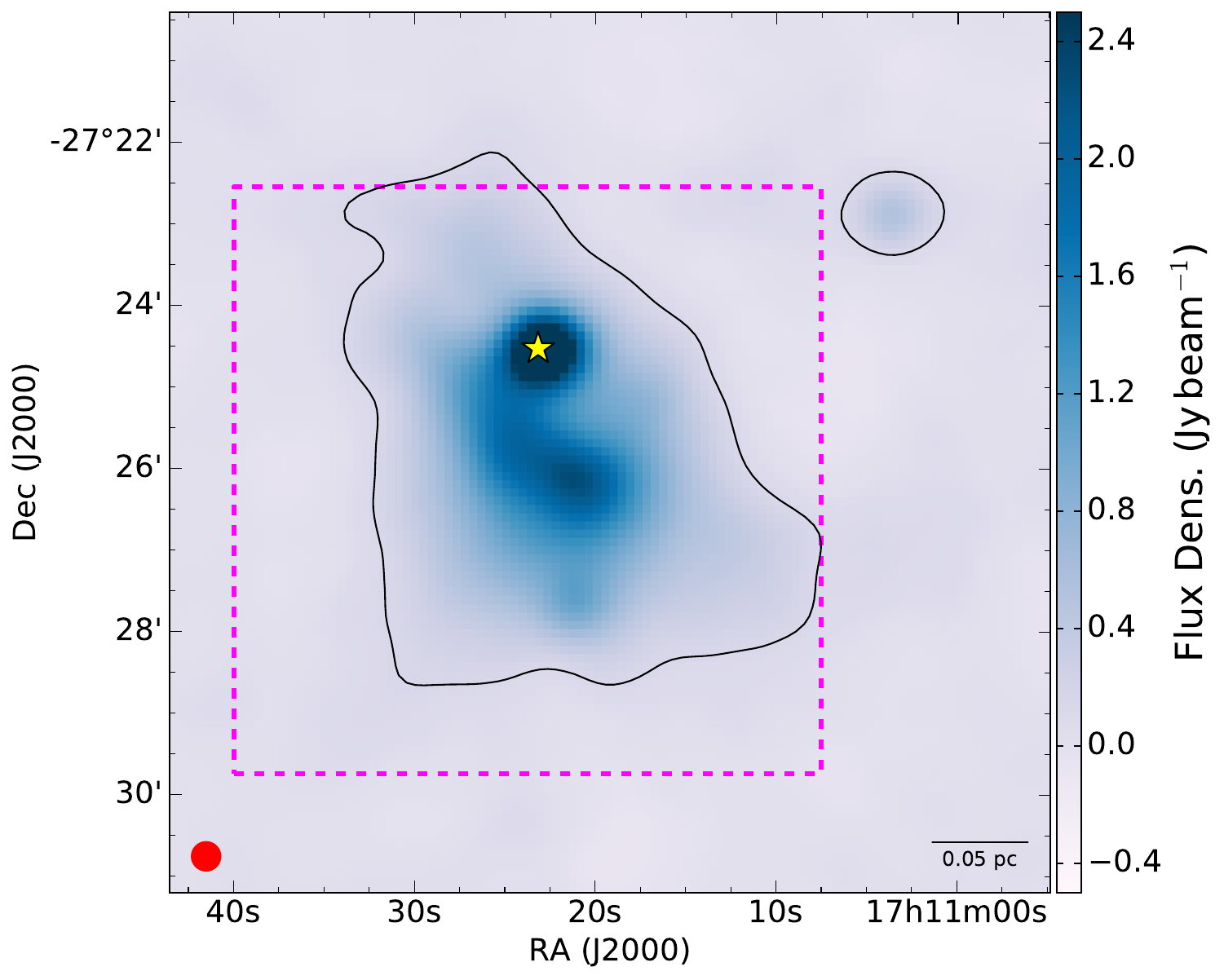}
   \caption{The APEX map showing thermal dust emission at $870\, \mu$m. The contour shows the $5\, \sigma$ level used in the mass computation ($\sigma = 0.026\,$Jy/beam). The bright region in the top-right corner, corresponding to the position of the YSO  BHB07-01 (B01), was excluded from the virial analysis. The yellow star shows the position of B11, and the magenta box shows the extension of ammonia data.}
  \label{APEX}
   \end{figure}

Figure \ref{APEX} shows the $870 \,\mu$m continuum image of B59 from APEX. The source mass has been calculated using these emission data at $870 \,\mu$m and:
\begin{equation}
M = f \frac{D^2 S_{\nu}}{B_{\nu}(\text{T}_{dust})  \kappa_{\nu}} \; ,
\end{equation}
where $f$ is the gas-to-dust ratio, assumed to be 100 \citep{Hildebrand83}, $D = (145 \pm 16)\,$pc is the source distance \citep{AlvesFranco07}, and $B_{\nu}(\text{T}_{dust})$ is the Planck function at a given dust temperature, which is assumed to be T$_{dust} = 15\,$K (consistent with the mean value found in the \tdust map, see Sec. \ref{H2Comp}). To be consistent with the Herschel data analysis (see Sec. \ref{CompHersch}), we have calculated the opacity at 870$\, \mu$m following: 
\begin{equation}
\kappa_{870} = \kappa_{250} \left ( \frac{250 \, \mu \text{m}}{870 \, \mu \text{m}}\right ) ^{\beta} = 1.54 \, \text{cm}^2 \, \text{g}^{-1} \; ,
\end{equation}
where the reference value at 250$\, \mu$m is $\kappa_{250} = 10.0\,$cm$^2$g$^{-1}$ \citep{Hildebrand83} and $\beta = 1.5$.  $S_{\nu}$ is the total flux at $870\, \mu$m derived from the 5$\, \sigma$ contour of the APEX map. In Figure \ref{APEX} is clearly visible the bright envelope of B11. 
Since we are interested in the cold core for the virial analysis, we fitted a bi-dimensional gaussian to the source to derive its integrated intensity flux. The calculation gives $S_{\text{B11}} = 15 \,$Jy, and we subtracted this value to the total flux $S_{\nu}$. The uncertainty on the latter has been estimated taking into account the calibration uncertainty ($10\%$, \citealt{Dumke10}). The computed total flux is $(111 \pm 11)\,$Jy and the obtained mass is $M = (24.2 \pm 5.9) \, M_{\odot}$. This value is in good agreement with previous estimations \citep{DuarteCabral12, RomanZuniga09,Rathborne09}. It is worth saying that if we apply the same approach to compute the mass from the Herschel fluxes, we obtain a result of $M \approx 10\,M_{\odot}$, a factor of 2 lower than our estimation with APEX. This difference can be due to multiple reasons, such as the dust opacity intrinsic uncertainty or a possibly too harsh background subtraction on the Herschel maps (which is though necessary to compare the ammonia results with the dust emission arising for the very same region). Moreover, at SPIRE wavelengths the Planck function is more sensible to the temperature that at APEX one. Given all of these, the discrepancy found seems reasonable, and the following analysis will be done with the APEX derived results. \par
The gravitational energy has been computed considering the source as a sphere, with a density distribution of the form $\rho \propto r^{-k}$:
\begin{equation}
U = - a \frac{3}{5} \frac{G M^2}{R},
\end{equation}
where $a = (1-k/3)/(1-2k/5)$, and we have set $k = 1.0$, a value that we later let vary. The radius $R$ corresponds to the effective radius of a circle with the same area of the region used to compute the total flux. We obtain $R = (0.120 \pm 0.013)\,$pc. The result for the the virial ratio is then$\alpha_{vir} = (0.43 \pm0.19)$, where the uncertainty has been calculated using standard error propagation. The B59 core is thus gravitationally virialized. The value of the virial parameter that we obtain is indeed low compared to values found in others studies of low-mass star-forming regions, according to which most of the cores are pressure confined (see for instance \citealt{Pattle15} or \citealt{Kirk17}). This confinement situation is true also in the Pipe, where the core population is mainly pressure confined as shown by \cite{Lada08}. As in other similar studies (i.e. \citealt{Kauffmann13}), these authors also indicated a trend between the virial parameter and the core mass: the higher the mass, the lower $\alpha_{vir}$. If we extrapolate their relation $\alpha_{vir} \propto M^{-0.66}$ at a mass of $25 \, M_{\odot}$ (approximately the mass of B59), we obtain $\alpha_{vir} \approx 0.80$, not far from the results in our analysis, once errors are taken into account. \par
Our results are robust against variations in the assumptions made for the various parameters. The dust spectral index $\beta$, for example, is usually taken in the range $[1.5,2.0]$ in star forming regions at core scales. An increment in its value corresponds to an increment in the computed dust mass, making the core even more sub-virial (for instance, $\beta = 1.7$ leads to $\alpha_{vir}=0.33$). The other parameter for which we make a strong assumption is the slope $k$ of the density profile, which can vary from $k = 0$ ($a = 1$, flat profile) to $k = 2.5$ (then $a \rightarrow \infty$). Taking as extremal cases $k = 0$ and $k = 2.0$ , the virial parameter values obtained are $\alpha_{vir}=0.47$ and $\alpha_{vir}=0.28$ respectively. Within these variations of the parameters, thus, the core is still gravitationally bound. \par
The virial analysis just described assumes that B59 is a single core. This assumption is supported by the investigation of \cite{RomanZuniga12}, who computed an extinction map across the source carefully subtracting the YSOs contribution. The resulting profile is monolithic, leading to the conclusion that B59 is not fragmented in multiple cores, and it should collapse as a whole structure. Nevertheless, to our knowledge in literature there has not been any detection of contraction of the core. Ammonia inversion transitions are not suitable to detect infall asymmetries, due to their low critical density and opacity. However, contraction motions within dense cores also produce line broadening at their very center (see e.g. \citealt{Caselli02}), and our ammonia spectra do not exhibit any sign of this broadening. If the core is not collapsing, a mechanism of support against gravity is needed, and it can possibly be found in the magnetic field. We can estimate the magnetic field strength $B$ needed to balance the gravitational pressure by including the magnetic energy term in the virial theorem:
\begin{equation}
\label{VirComp}
U + 2T + \Omega_B + \Omega_P= 0 \; ,
\end{equation}
where $\Omega_B = V \cdot B^2 / (2 \mu_0)$, where $V$ is the source volume (assuming spherical geometry: $V = 4\pi R^3 / 3$) and $\mu_0$ is the vacuum permeability. Here $\Omega_P$ is the external pressure term, which takes into account the confining pressure arising from the cloud weight ($P_w$) and the turbulence pressure exerted by the higher velocity dispersion and lower density medium that surrounds the core ($P_t$). Using the formalism introduced by \cite{Pattle15}, one has $\Omega_P = -4 \pi R^3(P_w+P_t)$. The turbulent term can be expressed has $P_t = \rho_t \sigma_t^2$, where $\rho_t$ and $\sigma_t$ are respectively the density and the velocity dispersion of the turbulent surrounding medium. The pressure due to the cloud weight, on the other hand, is $P_w = \pi G \overline{\Sigma} \Sigma_s (\mu m_{\text{H}})^2$, where $\overline{\Sigma}$ is the mean column density of the cloud and $\Sigma_s$ the column density at the source position (see Appendix A in \citealt{Kirk17}). For the present work we use the estimations for the two pressures made by \cite{Lada08}, according to which $P_w =  1.38 \cdot 10^{-12}\,$Pa and $P_t =6.90 \cdot 10^{-13}\, $Pa. Inverting Equation \ref{VirComp}, and using the expressions for $U$ and $T$ previously introduced, we obtain:
\begin{equation}
B = \sqrt{\frac{3 \mu_0}{2 \pi R^3} \left( |U| + |\Omega_P| - 2T \right)} =(140\pm65) \, \mu G  \; ,
\end{equation}
which has to be considered a lower limit, since it represents the minimum $B$ strength able to halt the gravitational contraction. The magnetic field strength of the Pipe Nebula at cloud scale (i.e. at a volume density $n \approx 10^3 \,$cm$^{-3}$) toward B59 is $\approx 17 \, \mu$G \citep{Alves08}. Given the peak H$_2$ column density and the source radius that we have derived in the spherical approach, we can estimate that our data are tracing a medium with an average volume density of $\approx  4 \cdot 10^{4} \,$cm$^{-3}$. \cite{Crutcher99} estimated that if  the magnetic field is unimportant in the collapse, $B \propto \rho^{2/3}$ while if it is dominant over gravity $B \propto \rho ^{1/2}$. Scaling the value found in \cite{Alves08} to our density with these two models, we obtain $B \approx 200 \, \mu$G and $B \approx 110 \, \mu$G respectively, in good agreement with our estimations. Given the uncertainties, we thus can not discriminate  between the two scenarios discussed by \cite{Crutcher99}.

 \section{Conclusions}
We have investigated the protostellar core B59 using primarily ammonia inversion transition spectra and ancillary dust continuum data, achieving a comprehensive view of the kinematics and dynamics of the core. The spectral fit performed on \am spectra reveals that B59 has a prominent cold, dense and velocity-coherent structure. The analysis of the turbulence level in the region shows that the non-thermal motions at the coherent core are subsonic, with a sharp transition to a supersonic regime in the surroundings. 
The coherent core is not clearly associated with any young stellar object identified in the c2d survey. In the northern part of the source, in correspondence with the position of B11 there is a change in the dynamical scenario, with higher velocity dispersion and temperature with respect to the quiescent and cold core. Here the kinetic temperatures are as high as $16-18\,$K, requiring possibly an extra source of heating in addition to the standard radiation field. This heating is most likely due to the internal radiation of B11, which is supported by the high levels of dust thermal emission from its dense envelope. B11 also powers a bipolar outflow likely responsible for the supersonic gas motions seen with the molecular data. The YSO envelope is deficient in ammonia, due to depletion produced by the release of CO from dust grains back to gas phase. \par
We have analyzed the velocity dispersion map and $\sigma_{NT}/C_S$ ratio to study the core kinematics. Sources B10 and B11 are associated with supersonic turbulent motions, suggesting that they are truly embedded within the source and are affecting through outflow feedback the surrounding gas. On the contrary, sources B04, B08, B14, B15 and B16 (which are all Class II objects) are found in quiescent regions, where the gas is subsonic. Most likely, these are foreground or background objects with respect to the material traced by ammonia. B03 and B09 presents \sig values close to the transonic limit, while the $\sigma_{NT}/C_S$ value (available only for the latter) is consistent with subsonic turbulent motions. Since these two YSOs are found close to the edge of our map's coverage, we do not to draw any conclusion about them. In the case of B12 as well our analysis is not conclusive, since with our smoothed angular resolution this source is within one beam of distance from B11. \par
The virial ratio obtained analyzing the dust emission is well below its critical value, indicating that the core is sub-virial and gravitationally bounded. This result suggests that the protostellar feedback affects the source only locally, and does not impact its global dynamics. Given such a low virial parameter, and since the extinction map of the core exhibit a monolithic profile \citep{RomanZuniga12}, Barnard 59 should be collapsing as a whole structure. Nevertheless, no signs of such a contraction are known to the present day. Our ammonia data do not show a line broadening at the center of the core, which is usually an indication of contracting motions. Further investigations of the source with other molecular tracers, more suitable for the detection of infall asymmetries (such as HCO$^+$ and isotopologues) will possibly confirm this hypothesis. Moreover, higher resolution VLA data will help us resolve possible substructures within the source, testing the hypothesis that Barnard 59 is a monolithic structure. \par
If the core is truly in equilibrium, since in our virial analysis we have taken into account only the internal pressure given by thermal and turbulent motions, the aforementioned considerations lead to the conclusion that another source of pressure is halting the gravitational collapse. Most likely, this source of support has to be the magnetic field, for which we have estimated a lower limit of $B \approx 140 \,\mu$G (see Sec. \ref{VirialAna}). At cloud scale, the Pipe Nebula is characterized by a strong magnetic field \citep{Alves08}. Future polarization observations of B59 will allow for an estimation of the magnetic pressure term and its role in the dynamical evolution of the core. Nevertheless, our data support the idea that Barnard 59 is not collapsing, and will not form new stars at least in the near future.

\acknowledgments
We thank Dr. Jorma Harju for his kind help in the analysis of Herschel data.


\appendix

\section{Determination of physical properties from NH$_3$ spectral fit \label{NH3app}}

The main assumption typically made in NH$_3$ line analysis is that \tex is the same for all of them \citep{HoTownes83, Rosolowsky08, Friesen09}. The system is thus characterized by three distinct temperatures:
\begin{itemize}
\item kinetic temperature T$_K$, which establishes the Maxwellian velocity distribution of the gas components;
\item excitation temperature T$_{ex}$, which regulates the population of upper ($u$) and lower ($l$) state in each $(J,K)$ rotational level following a Boltzmann distribution;
\item rotational temperature T$_{rot}$, according to which the different rotational levels are populated. Note that the terminology is not fully appropriate, since we are not analyzing purely rotational transitions.
\end{itemize}
\par
The approach consists of generating a model spectrum in main beam temperature and performing a non-linear regression to estimate best-fit parameter values and their uncertainties. The standard solution of the radiative transfer problem asserts that the T$_{MB}$ of a transition is:

\begin{equation}
\text{T}_{MB}=  A \, \left[ 1 - \exp \left(- \tau (v) \right) \right ] \; ,
\end{equation}
where the amplitude $A=\eta_f \left( J_{\nu}(T_{ex}) -  J_{\nu}(T_{b}) \right)$. $T_b$ is the background temperature , $\eta_f$ is the beam filling factor, and $ J_{\nu}(T)= (h \nu /k) \left(\exp (h \nu/ kT) -1 \right)^{-1}$.  Therefore, we express the optical depth $\tau(\nu)$ as a function of the input parameters, i.e., T$_{ex}$, T$_K$, $V_{lsr}$, $\sigma_V$, and N(NH$_3$). The absorption coefficient per unit of frequency can be written as:
\begin{equation}
\alpha_{\nu} = \frac{c^2}{8 \pi \nu_{rest}^2} \frac{g_u}{g_l}n_l A_{ul} \left[ 1 - \exp \left( - \frac{\Delta E}{k_B \text{\tex}} \right )\right] \, \phi (\nu) \; ,
\end{equation}
where $\phi (\nu)$ is the line profile function, which must be normalized to 1. Also, $\nu_{rest}$ is the line rest frequency  and $g_u$ and $g_l$ are the statistical weights for the upper and lower levels respectively, and the levels are separated in energy by $\Delta E$. In addition, $A_{ul}$ is the Einstein coefficient and $n_l$ is the lower state density. The latter is related to its upper counterpart $n_u$ through the Boltzmann equation:
\begin{equation}
\frac{n_{u}}{n_{l}}=\frac{N_{u}}{N_{l}} = \frac{g_{u}}{g_{l}} \exp \left(-\frac{\Delta E}{ k_B \text{\tex}} \right ) \; .
\label{boltzmann1}
\end{equation}
One must note that the first equivalence in \ref{boltzmann1} holds because we are assuming a homogeneous slab of material, and therefore volume density ratios are equal to column density ones.  For the (1,1) rotational state, $N_{11}$ = $N_u + N_l$. Thus one can easily obtain:
\begin{equation}
N_l = \frac{N_{11}}{1 + \frac{g_{u}}{g_{l}} \exp \left(-\frac{\Delta E}{k_B \text{\tex}} \right )} \; .
\end{equation}
$N_{11}$ is related to $N_{22}$ again according to the Boltzmann equation:
\begin{equation}
\frac{n_{22}}{n_{11}}=\frac{N_{22}}{N_{11}} = \frac{g_{22}}{g_{11}}  \left(-\frac{\Delta E_{12}}{k_B \text{T}_{rot}} \right ) \; , 
\label{boltzmann2}
\end{equation}
assuming that $\Delta E_{12} $ is the energy difference between the two levels. The total optical depth $\tau$ is obtained integrating $\alpha_{\nu}$ over frequency and line of sight:
\begin{eqnarray}
\tau  = \iint  \alpha_{\nu} \, d\nu ds  & = & \frac{c^2}{8 \pi \nu_{rest}^2} \frac{g_u}{g_l}N_l A_{ul} \left[ 1 - \exp \left( - \frac{\Delta E}{k_B \text{\tex}} \right )\right] \nonumber \\
& = & \frac{c^2}{8 \pi \nu_{rest}^2} \frac{g_u}{g_l}N_{11} A_{ul}  \frac{ 1 - \exp \left( - \frac{\Delta E}{k_B \text{\tex}} \right )}{1 + \frac{g_{u}}{g_{l}} \exp \left(-\frac{\Delta E}{k_B \text{\tex}} \right )} \; . 
\end{eqnarray}

The Einstein coefficient for dipole emission depends on the dipole matrix element $ \mu_{ul} $ \citep{MangumShirley15}:
\begin{equation}
A_{ul} = \frac{64 \pi^4 \nu_{rest}^3 }{3 h c g_u} \left| \mu_{ul} \right |^2 \; .
\end{equation}
To finally relate T$_{MB}$ and the total column density of the molecule, one can notice that $N_{11}$ is linked to \col through the rotational partition function $Z_{rot}$:
\begin{eqnarray}
&N_{11} = \text{\col} \frac{Z_{11}}{Z_{rot}} =  \\
&\text{\col} Z_{11}  \left \{ \sum_J (2J+1) \exp \left (-  \frac{h \left[ BJ(J+1) + (C-B) J^2\right]}{k_B \text{T}_{rot}}  \right)  \right \}^{-1} \; ,
\end{eqnarray}
where B and C are the rotational constants, equal to $298117 \,$MHz and $186726\,$MHz \citep{Pickett98}. We have considered para-states only. To obtain a complete model spectrum, we have to specify the line profile $\phi (\nu)$, which according to our hypothesis is a weighted sum of Gaussian lines, one for each hyperfine component. The weights correspond to the relative intensities and are derived from quantum mechanics calculations (see \citealt{MangumShirley15}). \par
The observed spectra can be fitted using the models produced with the above technique. A further clarification, however, is required: this approach only involves the rotational temperature, while we are mainly interested in the kinetic one. To obtain the latter, one can consider a three-state system composed by (1,1), (2,2) and (2,1) levels. The transition between different K-ladders is allowed only through collisions. Once the level (2,1) is populated, it rapidly decays radiatively to the (1,1). Thus, solving for the detailed balance taking into account spontaneous emission and collisional rates leads to the following equation \citep{Swift05}:
\begin{equation}
\text{T}_{rot} = \text{\tk} \left \{ 1 + \frac{\text{\tk}}{\text{T}_{12}} \ln \left [ 1+0.6 \exp(-15.7 \text{K} / \text{\tk})  \right ]   \right \}^{-1} \; ,
\end{equation}
where T$_{12}$ is the temperature corresponding to $\Delta E_{12} $ ( $=41.5\,$K).

\section{Herschel data analysis \label{Herschel}}
The dust emission can be modeled as grey-body emission, characterized by a frequency dependent optical depth $\tau_{\nu}$:
\begin{equation}
\label{graybody}
I_{\nu} = B_{\nu}(\text{T}) \left(  1- e^{-\tau_{\nu}} \right ) \; .
\end{equation}
The optical depth itself depends on two fundamental quantities: the opacity $\kappa_{\nu}$ and the mass density $\rho$ of the source:
\begin{equation}
\tau_{\nu} = \int_{LOS} \kappa_{\nu} \rho \, ds \; . 
\end{equation}
Given that the gas column density N(H$_2$) is the integration of the volume density $n(\text{H}_2)$ along the line of sight, and that $n(\text{H}_2) = \rho /(\mu_{\text{H}_2} m_{\text{H}})$ (where $\mu_{\text{H}_2} = 2.8$ is the mean molecular weight per hydrogen molecule), one obtains:
\begin{equation}
\label{taucol}
\text{N(H}_2) = \frac{\tau_{\nu}}{\mu_{\text{H}_2} m_{\text{H}} \kappa_{\nu} } \; ,
\end{equation}
Therefore, once the optical depth is known, one can infer the column density of the source, assuming that the opacity is known as well. For the latter, a power-law expression is usually adopted, making use of the dust opacity index $\beta$:
\begin{equation}
\kappa_{\nu} = \kappa_0 \left ( \frac{\nu}{\nu_0}\right ) ^{\beta} \; .
\end{equation}
We used $\beta_{\nu} = 1.5$, which seems suitable for star forming regions \citep{Walker90, Mezger90} and the dust opacity $\kappa_0 = 0.1\,$cm$^2$g$^{-1}$ \citep{Hildebrand83, Beckwith90} at the reference frequency corresponding to the wavelength $\lambda_0 = 250\,\mu$m. This value already includes a dust-to-mass ratio of 100. We also took into account the optically thin limit of Equation \ref{graybody}, which reads $I_{\nu} \approx B_{\nu}(\text{T}) \, \tau_{\nu}$. \par
We fit\footnote{See \cite{Hardegree-Ullman13} for details on the code.} the three photometric points from SPIRE data, taking into account the color corrections for grey-body and extended sources, and we obtain a map of \tdust and optical depth. The latter is converted into a column density map according to Equation \ref{taucol}. Uncertainties are evaluated through a Monte Carlo method, iterating the described procedure 1000 times. At each iteration a random noise is added to the SPIRE maps, taking into account both calibration uncertainties ($7\%$) and the fluctuations that the detector response can present. See SPIRE HandBook (Version 3.1) for details about color corrections or flux uncertainties.


\begin{thebibliography}{100}

\bibitem[Alves et al.(2008)]{Alves08} Alves, F.~O., Franco, G.~A.~P., \& Girart, J.~M.\ 2008, \aap, 486, L13 

\bibitem[Alves \& Franco(2007)]{AlvesFranco07} Alves, F.~O., \& Franco, G.~A.~P.\ 2007, \aap, 470, 597 

\bibitem[Andre et al.(2000)]{Andre00} Andre, P., Ward-Thompson, D., \& Barsony, M.\ 2000, Protostars and Planets IV, 59 

\bibitem[Andr{\'e} et al.(2010)]{Andre10} Andr{\'e}, P., Men'shchikov, A., Bontemps, S., et al.\ 2010, \aap, 518, L102 

\bibitem[Beckwith et al.(1990)]{Beckwith90} Beckwith, S.~V.~W., Sargent, A.~I., Chini, R.~S., \& Guesten, R.\ 1990, \aj, 99, 924 

\bibitem[Benson \& Myers(1989)]{BensonMyers89} Benson, P.~J., \& Myers, P.~C.\ 1989, \apjs, 71, 89 

\bibitem[Bertoldi \& McKee(1992)]{Bertoldi92} Bertoldi, F., \& McKee, C.~F.\ 1992, \apj, 395, 140 

\bibitem[Bohlin et al.(1978)]{Bohlin78} Bohlin, R.~C., Savage, B.~D., \& Drake, J.~F.\ 1978, \apj, 224, 132 

\bibitem[Brooke et al.(2007)]{Brooke07} Brooke, T.~Y., Huard, T.~L., Bourke, T.~L., et al.\ 2007, \apj, 655, 364 

\bibitem[Caselli et al.(2002)]{Caselli02} Caselli, P., Benson, P.~J., Myers, P.~C., \& Tafalla, M.\ 2002, \apj, 572, 238 

\bibitem[Charnley \& Rodgers(2002)]{CharnleyRodgers02} Charnley, S.~B., \& Rodgers, S.~D.\ 2002, \apjl, 569, L133 

\bibitem[Crutcher(1999)]{Crutcher99} Crutcher, R.~M.\ 1999, \apj, 520, 706 

\bibitem[Duarte-Cabral et al.(2012)]{DuarteCabral12} Duarte-Cabral, A., Chrysostomou, A., Peretto, N., et al.\ 2012, \aap, 543, A140 

\bibitem[Dumke \& Mac-Auliffe(2010)]{Dumke10} Dumke, M., \& Mac-Auliffe, F.\ 2010, \procspie, 7737, 77371J 


\bibitem[Elmegreen \& Scalo(2004)]{Elmegreen04} Elmegreen, B.~G., \& Scalo, J.\ 2004, \araa, 42, 211 

\bibitem[Evans et al.(2009)]{Evans09} Evans, N.~J., II, Dunham, M.~M., J{\o}rgensen, J.~K., et al.\ 2009, \apjs, 181, 321-350 

\bibitem[Evans et al.(2001)]{Evans01} Evans, N.~J., II, Rawlings, J.~M.~C., Shirley, Y.~L., \& Mundy, L.~G.\ 2001, \apj, 557, 193 

\bibitem[Forbrich et al.(2009)]{Forbrich09} Forbrich, J., Lada, C.~J., Muench, A.~A., Alves, J., \& Lombardi, M.\ 2009, \apj, 704, 292 

\bibitem[Foster et al.(2009)]{Foster09} Foster, J.~B., Rosolowsky, E.~W., Kauffmann, J., et al.\ 2009, \apj, 696, 298 

\bibitem[Franco et al.(2010)]{Franco10} Franco, G.~A.~P., Alves, F.~O., \& Girart, J.~M.\ 2010, \apj, 723, 146 

\bibitem[Friesen et al.(2009)]{Friesen09} Friesen, R.~K., Di Francesco, J., Shirley, Y.~L., \& Myers, P.~C.\ 2009, \apj, 697, 1457 

\bibitem[Friesen et al.(2017)]{FriesenPineda17} Friesen, R.~K., Pineda, J.~E., co-PIs, et al.\ 2017, \apj, 843, 63 

\bibitem[Ginsburg \& Mirocha(2011)]{Ginsburg11} Ginsburg, A., \& Mirocha, J.\ 2011, Astrophysics Source Code Library, ascl:1109.001 

\bibitem[Goodman et al.(1998)]{Goodman98} Goodman, A.~A., Barranco, J.~A., Wilner, D.~J., \& Heyer, M.~H.\ 1998, \apj, 504, 223 

\bibitem[Hara et al.(2013)]{Hara13} Hara, C., Shimajiri, Y., Tsukagoshi, T., et al.\ 2013, \apj, 771, 128 

\bibitem[Hardegree-Ullman et al.(2013)]{Hardegree-Ullman13} Hardegree-Ullman, E., Harju, J., Juvela, M., et al.\ 2013, \apj, 763, 45 

\bibitem[Hildebrand(1983)]{Hildebrand83} Hildebrand, R.~H.\ 1983, \qjras, 24, 267 

\bibitem[Ho \& Townes(1983)]{HoTownes83} Ho, P.~T.~P., \& Townes, C.~H.\ 1983, \araa, 21, 239 

\bibitem[Kauffmann et al.(2013)]{Kauffmann13} Kauffmann, J., Pillai, T., \& Goldsmith, P.~F.\ 2013, \apj, 779, 185 

\bibitem[Kirk et al.(2017)]{Kirk17} Kirk, H., Friesen, R.~K., Pineda, J.~E., et al.\ 2017, \apj, 846, 144 

\bibitem[Lada et al.(2008)]{Lada08} Lada, C.~J., Muench, A.~A., Rathborne, J., Alves, J.~F., \& Lombardi, M.\ 2008, \apj, 672, 410-422 

\bibitem[Lada \& Wilking(1984)]{Lada84} Lada, C.~J., \& Wilking, B.~A.\ 1984, \apj, 287, 610

\bibitem[Lombardi et al.(2006)]{Lombardi06} Lombardi, M., Alves, J., \& Lada, C.~J.\ 2006, \aap, 454, 781 

\bibitem[Mac Low \& Klessen(2004)]{MacLow04} Mac Low, M.-M., \& Klessen, R.~S.\ 2004, Reviews of Modern Physics, 76, 125 

\bibitem[Mangum et al.(2007)]{Mangum07} Mangum, J.~G., Emerson, D.~T., \& Greisen, E.~W.\ 2007, \aap, 474, 679 

\bibitem[Mangum \& Shirley(2015)]{MangumShirley15} Mangum, J.~G., \& Shirley, Y.~L.\ 2015, \pasp, 127, 266 

\bibitem[Masters et al.(2011)]{Masters11} Masters, J., Garwood, B., Langston, G., \& Shelton, A.\ 2011, Astronomical Data Analysis Software and Systems XX, 442, 127 

\bibitem[Mezger et al.(1990)]{Mezger90} Mezger, P.~G., Zylka, R., \& Wink, J.~E.\ 1990, \aap, 228, 95 

\bibitem[Myers et al.(1991)]{Myers91} Myers, P.~C., Ladd, E.~F., \& Fuller, G.~A.\ 1991, \apjl, 372, L95 

\bibitem[Nakamura \& Li(2008)]{Nakamura08} Nakamura, F., \& Li, Z.-Y.\ 2008, \apj, 687, 354-375

\bibitem[Offner \& Arce(2014)]{Offner14} Offner, S.~S.~R., \& Arce, H.~G.\ 2014, \apj, 784, 61 

\bibitem[Onishi et al.(1999)]{Onishi99} Onishi, T., Kawamura, A., Abe, R., et al.\ 1999, \pasj, 51, 871 

\bibitem[Palmeirim et al.(2013)]{Palmeirim13} Palmeirim, P., Andr{\'e}, P., Kirk, J., et al.\ 2013, \aap, 550, A38 

\bibitem[Pattle et al.(2015)]{Pattle15} Pattle, K., Ward-Thompson, D., Kirk, J.~M., et al.\ 2015, \mnras, 450, 1094 

\bibitem[Peretto et al.(2012)]{Peretto12} Peretto, N., Andr{\'e}, P., K{\"o}nyves, V., et al.\ 2012, \aap, 541, A63 

\bibitem[Pickett et al.(1998)]{Pickett98} Pickett, H.~M., Poynter, R.~L., Cohen, E.~A., et al.\ 1998, \jqsrt, 60, 883 

\bibitem[Pineda et al.(2008)]{Pineda08} Pineda, J.~E., Caselli, P., \& Goodman, A.~A.\ 2008, \apj, 679, 481-496 

\bibitem[Pineda et al.(2010)]{Pineda10} Pineda, J.~E., Goodman, A.~A., Arce, H.~G., et al.\ 2010, \apjl, 712, L116 

\bibitem[Pineda et al.(2015)]{Pineda15} Pineda, J.~E., Offner, S.~S.~R., Parker, R.~J., et al.\ 2015, \nat, 518, 213 

\bibitem[Planck Collaboration et al.(2016)]{PlanckXXXV} Planck Collaboration, Ade, P.~A.~R., Aghanim, N., et al.\ 2016, \aap, 586, A138 

\bibitem[Rathborne et al.(2009)]{Rathborne09} Rathborne, J.~M., Lada, C.~J., Muench, A.~A., et al.\ 2009, \apj, 699, 742 

\bibitem[Rodgers \& Charnley(2008)]{RodgersCharnley08} Rodgers, S.~D., \& Charnley, S.~B.\ 2008, \apj, 689, 1448-1455 

\bibitem[Rom{\'a}n-Z{\'u}{\~n}iga et al.(2009)]{RomanZuniga09} Rom{\'a}n-Z{\'u}{\~n}iga, C.~G., Lada, C.~J., \& Alves, J.~F.\ 2009, \apj, 704, 183 

\bibitem[Rom{\'a}n-Z{\'u}{\~n}iga et al.(2012)]{RomanZuniga12} Rom{\'a}n-Z{\'u}{\~n}iga, C.~G., Frau, P., Girart, J.~M., \& Alves, J.~F.\ 2012, \apj, 747, 149 

\bibitem[Rosolowsky et al.(2008)]{Rosolowsky08} Rosolowsky, E.~W., Pineda, J.~E., Foster, J.~B., et al.\ 2008, \apjs, 175, 509-521 

\bibitem[Seifried \& Walch(2015)]{Seifried15} Seifried, D., \& Walch, S.\ 2015, \mnras, 452, 2410 

\bibitem[Shirley(2015)]{Shirley15} Shirley, Y.~L.\ 2015, \pasp, 127, 299 

\bibitem[Swift et al.(2005)]{Swift05} Swift, J.~J., Welch, W.~J., \& Di Francesco, J.\ 2005, \apj, 620, 823 

\bibitem[Tafalla et al.(2002)]{Tafalla02} Tafalla, M., Myers, P.~C., Caselli, P., Walmsley, C.~M., \& Comito, C.\ 2002, \apj, 569, 815 

\bibitem[Tielens(2005)]{Tielens05} Tielens, A.~G.~G.~M.\ 2005, The Physics and Chemistry of the Interstellar Medium, by A.~G.~G.~M.~Tielens, pp.~.~ISBN 0521826349.~Cambridge, UK: Cambridge University Press,  2005., 

\bibitem[Walker et al.(1990)]{Walker90} Walker, C.~K., Adams, F.~C., \& Lada, C.~J.\ 1990, \apj, 349, 515 

\bibitem[Zucconi et al.(2001)]{Zucconi01} Zucconi, A., Walmsley, C.~M., \& Galli, D.\ 2001, \aap, 376, 650 


\end{thebibliography}

\end{document}